\documentclass[twocolumn,english,aps,english,showpacs,preprintnumbers,groupedaddress,amsmath,amssymb,pra]{revtex4}
\usepackage[T1]{fontenc}
\usepackage[latin9]{inputenc}
\usepackage{amsmath}
\usepackage{amssymb}
\usepackage{graphicx}
\usepackage{esint}

\makeatletter

\newcommand{\noun}[1]{\textsc{#1}}

\@ifundefined{textcolor}{}
{%
 \definecolor{BLACK}{gray}{0}
 \definecolor{WHITE}{gray}{1}
 \definecolor{RED}{rgb}{1,0,0}
 \definecolor{GREEN}{rgb}{0,1,0}
 \definecolor{BLUE}{rgb}{0,0,1}
 \definecolor{CYAN}{cmyk}{1,0,0,0}
 \definecolor{MAGENTA}{cmyk}{0,1,0,0}
 \definecolor{YELLOW}{cmyk}{0,0,1,0}
}

\usepackage{babel}

\makeatother

\usepackage{babel}

\begin{document}

\title{Brillouin precursors in Debye media}

\author{Bruno Macke}

\author{Bernard S\'{e}gard}

\email{bernard.segard@univ-lille-1.fr}

\affiliation{Laboratoire de Physique des Lasers, Atomes et Mol\'{e}cules , CNRS et
Universit\'{e} de Lille, 59655 Villeneuve d'Ascq, France}

\date{\today}
\begin{abstract}
We theoretically study the formation of Brillouin precursors in Debye
media. We point out that the precursors are visible only at propagation
distances such that the impulse response of the medium is essentially
determined by the frequency dependence of its absorption and is practically
Gaussian. By simple convolution, we then obtain explicit analytical
expressions of the transmitted waves generated by reference incident
waves, distinguishing precursor and main signal by a simple examination
of the long-time behavior of the overall signal. These expressions
are in good agreement with the signals obtained in numerical or real
experiments performed on water in the radio-frequency domain and explain
in particular some observed shapes of the precursor. Results are obtained for other remarkable incident waves. In addition, we show quite generally
that the shape of the Brillouin precursor appearing alone at sufficiently large propagation distance and the law giving its amplitude as a function of this distance do not depend on the precise form of the
incident wave but only on its integral properties. The incidence of
a static conductivity of the medium is also examined and explicit
analytical results are again given in the limit of weak and strong
conductivities. 
\end{abstract}

\pacs{42.50.Md, 41.20.Jb, 42.25.Bs}

\maketitle

\section{INTRODUCTION\label{sec:INTRODUCTION}}

In their celebrated papers \cite{som14,bri14,bri32,bri60} on the
propagation of a step-modulated optical wave through a single-resonance
Lorentz medium, Sommerfeld and Brillouin found that, in suitable conditions,
the arrival of the signal at the carrier frequency (``main signal\textquotedblright )
is preceded by that of a first and a second transient, respectively
generated from the high and low frequencies contained in the spectrum
of the incident wave. More than a century after their discovery, these
transients, currently named Sommerfeld and Brillouin precursors, continue
to raise considerable interest, due in particular to their sub-exponential
attenuation with the propagation distance. An abundant bibliography
on the subject is given in  Ref \cite{ou09} and more recent related studies
are reported in Refs \cite{je09,ou10,bm11,cia11,bm12}.

In fact the simultaneous observation of well-distinguishable Sommerfeld
and Brillouin precursors, as considered by their discoverers, requires
experimental conditions that seem unrealizable in optics \cite{bm11}.
A qualitative demonstration of separated Sommerfeld and Brillouin
precursors has only been performed in microwaves by using guiding structures
with dispersion characteristics similar to those of a Lorentz medium
\cite{pl69}. In optics, only the unique precursor resulting from
a complete overlapping of the Sommerfeld and Brillouin precursors
\cite{aa88,bm13} has been actually observed \cite{bs87,aa91,je06,wei09,bm10,zh13}.

Precursors are obviously not specific to electromagnetic waves and
Lorentz media. A convincing demonstration of Sommerfeld precursors
has been performed by using elastic waves propagating on a liquid
surface \cite{fa03}. We consider here electromagnetic waves propagating
in dielectric media whose susceptibility is determined by the partial
orientation of molecular dipoles under the effect of the applied electric
field \cite{de30}. These currently called Debye media are opaque
(transparent) at high (low) frequency. Only the Brillouin precursor
is thus expected to be observable in good conditions, without the
overlapping problem encountered in Lorentz media. Propagation of waves
in Debye media has been extensively studied in the past, with particular
attention paid to the important case of water in the radio-frequency
and microwave domains. For papers related to the present study, see,
e.g., Refs \cite{al89,ro96,fa97,ka98,li01,sto01,ro04,ou05,pi09,sa09,da10,ca11,bm12,ou12}.
Albanese \emph{et al.} were the first to refer to the Brillouin precursors
in Debye media \cite{al89}. They \emph{numerically} studied the propagation
in water of sine waves at 1 GHz with square or trapezoidal envelopes
and noticed ``a well-formed transient field that appears similar
to the Brillouin precursor observed in media with anomalous dispersion\textquotedblright . Stoudt \emph{et al}. \cite{sto01} performed the corresponding experiments,
achieving \emph{direct detection of Brillouin precursors}.
Indirect experimental demonstrations were achieved later \cite{pi09,da10}.
Oughstun \emph{et al}. theoretically studied at length the propagation
of waves with a step, rectangular or trapezoidal envelope in Debye
media by combining numerical computations and analytical calculations
using saddle-point methods \cite{ou05,ou09,ou12}. Important features
of the transmitted wave and, in particular, some remarkable shapes
of the Brillouin precursor were unfortunately overlooked in these
works.

In the present paper, the problem is greatly simplified by remarking
that the precursor is only visible (discernible from the main signal)
at propagation distances for which the impulse response of the medium
is practically Gaussian. By convolution procedures, this enables us
to obtain \emph{fully analytical} expressions of the transmitted waves
generated by different incident waves, distinguishing main signal
and precursor by a simple examination of the long-time behavior
of the overall signal. The arrangement of our paper is as follows.
In Sec \ref{sec:deltafunction}, we present our approximation
and give the corresponding impulse response of the medium. We derive
in Sec \ref{sec:StepSin} the transmitted wave generated from
incident sine waves with a step or rectangular envelope. The response
to sine-waves with linearly varying amplitude is studied in Sec \ref{sec:Trapeze}. General properties of the precursors in the strict asymptotic limit are established in Sec \ref{sec:AsymptoticProperties}
and the effects of a static conductivity of the medium are examined
in Sec \ref{sec:Conduction}. We finally conclude in Sec \ref{sec:CONCLUSION}
by summarizing our main results.

\section{IMPULSE RESPONSE OF THE MEDIUM\label{sec:deltafunction}}

As in the experiments on water reported in Refs \cite{fa97,sto01}, we
consider transverse electromagnetic waves propagating in a coaxial
transmission line containing the Debye medium. This coaxial geometry
has the advantage of having a ``flat frequency response down to and
including DC\textquotedblright{} \cite{sto01}, as required for a
good observation of the Brillouin precursors. We denote $\ell$ as the
length of the transmission line, $v(0,t)$ {[}$v(\ell,t)${]} as the
voltage at its input {[}output{]} inside the medium and $V(0,\omega)$
{[}$V(\ell,\omega)${]} as its Laplace-Fourier transform. In the frequency
domain the medium is fully characterized by its transfer function
$H(\ell,\omega)$ relating $V(\ell,\omega)$ to $V(0,\omega)$ \cite{pap87}:

. 
\begin{equation}
V(\ell,\omega)=H(\ell,\omega)V(0,\omega)\label{eq:un}
\end{equation}
with 
\begin{equation}
H(\ell,\omega)=\exp\left[-i\frac{\omega\ell}{c}\tilde{n}(\omega)\right]\label{eq:deux}
\end{equation}
where $\omega$, $c$ and $\tilde{n}(\omega)$ are the (angular) frequency,
the light velocity in vacuum and the complex refractive index of the
medium, respectively. For Debye media, we have 
\begin{equation}
\tilde{n}(\omega)=\sqrt{n_{\infty}^{2}+\frac{n_{0}^{2}-n_{\infty}^{2}}{1+i\omega\tau}}\label{eq:trois}
\end{equation}
In this expression $n_{\infty}$, $n_{0}$ and $\tau$ respectively
denote the refractive index at high frequency (Debye plateau), the
refractive index at vanishing frequency and the orientation relaxation
time of the polar molecules. Eq.(\ref{eq:trois}) provides a good
approximation of the refractive index of Debye media in the radiofrequency
and microwave domains. For deionized water, often taken as reference,
$n_{0}^{2}\approx79$ , $n_{\infty}^{2}\approx5.5$ and $\tau\approx8.5\,\mathrm{ps}$
may be considered as typical values \cite{sto01} and are used
in our calculations. For the frequencies up to 100 GHz
the real and imaginary parts of the refractive index derived from
Eq. (\ref{eq:trois}) with these parameters fit very well the measurements
reported in Ref \cite{se81}.

In the time domain, the medium will be characterized by its impulse
response $h(\ell,t)$, which is the inverse Fourier transform of $H(\ell,\omega)$,
and the output voltage is given by the convolution product

\begin{equation}
v(\ell,t)=h(\ell,t)\otimes v(0,t)\label{eq:cinq}
\end{equation}
$h(\ell,t)$ has no exact analytical form. Some general properties
of $h(\ell,t)$ can however be derived from $H(\ell,\omega)$. The
relation $\intop_{-\infty}^{+\infty}h(\ell,t)\,dt=H(\ell,0)=1$ shows
that it has a unit area. The location of its center of gravity $t_{B}=\left(\intop_{-\infty}^{+\infty}t\,h(\ell,t)\,dt\right)/\left(\intop_{-\infty}^{+\infty}h(\ell,t)\,dt\right)$,
its centered second moment or variance $\sigma^{2}=\left(\intop_{-\infty}^{+\infty}(t-t_{B})^{2}\,h(\ell,t)\,dt\right)/\left(\intop_{-\infty}^{+\infty}h(\ell,t)\,dt\right)$
and its centered third moment $\mu_{3}=\left(\intop_{-\infty}^{+\infty}(t-t_{B})^{3}\,h(\ell,t)\,dt\right)/\left(\intop_{-\infty}^{+\infty}h(\ell,t)\,dt\right)$
are respectively equal to the cumulants $k_{1}$, $k_{2}$ and $k_{3}$
of the transfer function \cite{bu04} as defined in the following
expansion: 
\begin{equation}
H(\ell,\omega)=\exp\left(\sum_{n=1}^{\infty}\frac{\left(-i\omega\right)^{n}}{n!}k_{n}\left(\ell\right)\right)\label{eq:six}
\end{equation}
In the case of a Debye medium, we deduce from Eqs.(\ref{eq:deux}) and (\ref{eq:trois})
$t_{B}=n_{0}\ell/c$ , $\sigma^{2}=\frac{n_{0}^{2}-n_{\infty}^{2}}{n_{0}}\left(\frac{\ell\tau}{c}\right)$
and $\mu_{3}=\frac{3\left(n_{0}^{2}-n_{\infty}^{2}\right)\left(3n_{0}^{2}+n_{\infty}^{2}\right)}{4n_{0}^{3}}\left(\frac{\ell\tau^{2}}{c}\right)$.
These expressions show that the center of gravity of $h(\ell,t)$
propagates at the phase velocity at zero frequency (equal to the group
velocity for this frequency) and has a root-mean-square duration $\sigma$
proportional to $\sqrt{\ell}$ and a positive skewness or asymmetry
$\xi=\tfrac{\mu_{3}}{\sigma^{3}}=\frac{3\left(3n_{0}^{2}+n_{\infty}^{2}\right)}{4\sqrt{n_{0}^{3}\left(n_{0}^{2}-n_{\infty}^{2}\right)}}\sqrt{\frac{c\tau}{\ell}}$.

\begin{figure}[h]
\begin{centering}
\includegraphics[width=85mm]{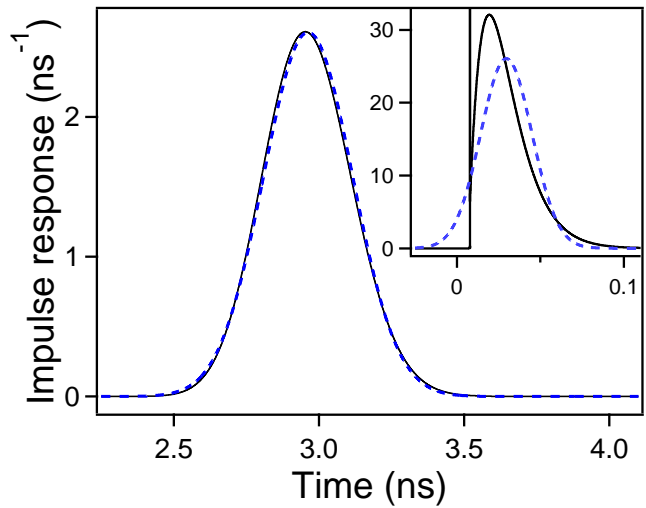} 
\par\end{centering}
\protect\caption{(Color online) Impulse response of deionized water as a function of
time for a propagation distance $\ell=10\:\mathrm{cm}$ leading to
$t_{B}=2.96\:\mathrm{ns}$ and $\beta=4.63\:\mathrm{\mathrm{n}s^{-1}}$.
The solid {[}dashed{]} line is the exact numerical result obtained
by FFT {[}the Gaussian shape given by Eq. (\ref{eq:huit}){]}. Inset: the same for $\ell=1\:\mathrm{mm}$,
leading to $t_{B}=29.6\:\mathrm{ps}$ and $\beta=46.3\:\mathrm{\mathrm{n}s^{-1}}$.\label{fig:DeltaResponse}}
\end{figure}

The previous results are valid for arbitrary propagation distances.
The expression of the skewness ($\xi\varpropto1/\sqrt{\ell}$) suggests
that the expansion of Eq. (\ref{eq:six}) may be limited to the term
$n=2$ when $\ell$ is large enough. Taking the origin of time at
$t=t_{B}$, the transfer function and the impulse response are reduced
to the Gaussians 
\begin{equation}
H(\ell,\omega)=e^{-\omega^{2}/(4\beta^{2})}=e^{-\alpha(\omega)\ell}\label{eq:sept}
\end{equation}
\begin{equation}
h(\ell,t)=\frac{\beta}{\sqrt{\pi}} e^{-\beta^{2}t^{2}}\label{eq:huit}
\end{equation}
where $\alpha(\omega)$ is the absorption coefficient of the medium
at the frequency $\omega$ and 
\begin{equation}
\beta=\frac{1}{\sigma\sqrt{2}}=\sqrt{\frac{cn_{0}}{2\ell\tau\left(n_{0}^{2}-n_{\infty}^{2}\right)}}\label{eq:neuf}
\end{equation}
$h(\ell,t)$ has then a peak amplitude (a duration) proportional to
$1/\sqrt{\ell}$ ($\sqrt{\ell}$) with a unit area, as expected. It
meets the principle of relativistic causality \cite{re0} as long
as $\exp\left[-\beta^{2}\left(n_{0}-n_{\infty}\right)^{2}\left(\ell/c\right)^{2}\right]$
is negligible, a condition superabundantly satisfied for the propagation
distances at which the Brillouin precursor is visible. Decomposing
the medium in $m$ subsections of impulse response $h(\ell/m,t)$,
the Gaussian form of $h(\ell,t)$ given by Eq. (\ref{eq:huit}) may
be considered as a consequence of the central limit theorem in a deterministic
case \cite{pap87}. It is also the limit when $\beta t_{B}\gg1$ of
the expressions of $h(\ell,t)$ obtained by direct studies in the
time-domain \cite{ro96,ka98,ro04}. When $\xi\ll1$, the condition
$\beta t_{B}\gg1$ is automatically fulfilled and the Gaussian form
$h(\ell,t)$ given by Eq. (\ref{eq:huit}) is expected to hold. This
is illustrated Fig \ref{fig:DeltaResponse}, obtained for $\ell=10\,\mathrm{cm}$
in deionized water. 
We have then $\beta t_{B}\approx13.7$ and the result given by Eq. (\ref{eq:huit})
is actually very close to the exact result, numerically derived from
the exact transfer function by fast Fourier transform (FFT). The main
effect of the residual skewness is a very slight reduction of the
time of the maximum by about $9.5\,\mathrm{ps}$ whereas $t_{B}\approx2.96\,\mathrm{ns}$.
For $\ell=1\:\mathrm{cm}$ ($\beta t_{B}\approx4.3$), the skewness
increases and the rise of $h(\ell,t)$ becomes significantly steeper
than its fall. However the Gaussian form of Eq. (\ref{eq:huit}) remains
a reasonable approximation of the exact result. This approximation
completely fails when $\beta t_{B}\lesssim1$ . The impulse response then
begins at the time $n_{\infty}\ell/c$ \cite{re0} by a Dirac peak
and a discontinuity. The weight $w$ of the former and the amplitude
$\Delta h$ of the latter are easily determined from an asymptotic
expansion of $H(\ell,\omega)$. We get $w=\exp\left[-\left(n_{0}^{2}-n_{\infty}^{2}\right)\ell/\left(2n_{\infty}c\tau\right)\right]$
and $\Delta h=w\ell\left(n_{0}^{2}-n_{\infty}^{2}\right)\left(n_{0}^{2}+3n_{\infty}^{2}\right)/\left(8c\tau^{2}n_{\infty}^{3}\right)$.
The inset of Fig. \ref{fig:DeltaResponse} shows the impulse response
obtained for $\ell=1\:\mathrm{mm}$ ($\beta t_{B}\approx1.37$). We
have then $w\approx2.14\times10^{-3}$ and $\Delta h\approx6.7\,\mathrm{ns^{-1}}$.
The previous results are only given for completeness. Indeed, it will
be shown later that the Brillouin precursor emerges from the main
signal for $\ell>20\:\mathrm{cm}$ ($\beta t_{B}>19.4$ ) and the
Gaussian approximation of $h(\ell,t)$ is then excellent. Note that
this remarkable form of the impulse response is not specific to the
Debye medium but holds at large enough propagation distances whenever
the transfer function can be expanded in cumulants. It applies in
particular to polar media with relaxation mechanisms more complex
than those considered here. We incidentally note that the addition
of a second relaxation time in the Debye model as considered in Refs \cite{ou05,ou09}
does not significantly modify the parameters $t_{B}$ and $\beta$.

The considerable advantage of the Gaussian impulse response is that
it can be convoluted with a great number of input signals to provide
exact analytical forms of the output signal. This is the case, e.g.,
when $v(0,t)$ is a step, an algebraic function, an exponential, a
Gaussian or an error function, when these functions modulate a sine
wave (eventually linearly chirped) or when $v(0,t)$ is a combination
of the previous signals \cite{al89,li01,sto01,ou05,ou09,da10,ou12,bm12}.
The relation $V(\ell,0)=H(\ell,0)V(0,0)$ quite generally shows that
the area of the output signal will be always equal to that of the
input signal. We also remark that the Gaussian impulse response is
obtained by neglecting the effects of the group-velocity dispersion.
This explains why the shape of the output signals observed in the
experiments reported in Ref \cite{sto01} was well reproduced by numerical
simulations taking only into account the frequency-dependence of the
medium absorption. This also means that the Brillouin precursors actually
observed or observable in Debye media essentially originate from the
latter and that group-velocity dispersion plays no significant role
in their formation.

In the previous theoretical analysis it is assumed as usual that the
input and output signals are measured inside the Debye medium. This
is generally not the case in the experiments. $H(\ell,\omega)$ should
then be multiplied by the transfer function $T(\omega)$ taking into
account the losses at the air-medium and medium-air interfaces, that
is 
\begin{equation}
T(\omega)=\left[\frac{2}{1+\widetilde{n}(\omega)}\right]\times\left[\frac{2}{1+1/\widetilde{n}(\omega)}\right]=\frac{4\widetilde{n}(\omega)}{\left[1+\widetilde{n}(\omega)\right]^{2}}\label{eq:onze}
\end{equation}
The main effect of $T(\omega)$ will be an overall reduction of the
amplitude of the output voltage by a factor $1/T(0)=\left(n_{0}+1\right)^{2}/\left(4n_{0}\right)$
that is $2.75$ for deionized water. The frequency-dependent effects
can be determined by expanding $T(\omega)/T(0)$ in cumulants as made
for $H(\ell,\omega)$ and exploiting the additivity property of the
cumulants. Again for deionized water, we find that $t_{B}$ ($\sigma^{2}$)
is reduced by about $3.15\,\mathrm{ps}$ ($15.7\,\mathrm{ps^{2}}$).
For $\ell\geq10\,\mathrm{cm}$, these quantities are fully negligible
compared to those associated with $H(\ell,\omega)$. The output signals
determined from $H(\ell,\omega)$ in the following should thus be
simply multiplied by $T(0)=4n_{0}/\left(1+n_{0}\right)^{2}$ and delayed
by the extra transit times in air when the voltages are measured outside
the medium.

\section{RESPONSE TO STEP-MODULATED SINE-WAVES \label{sec:StepSin} }

We consider in this section input signals of the form $u_{H}(t)\sin\left(\omega_{c}t\right)$
or $u_{H}(t)\cos\left(\omega_{c}t\right)$, where $u_{H}(t)$ is the
Heaviside unit step function. They consist of a sine wave of angular
frequency $\omega_{c}$ (the ``carrier\textquotedblright ) switched
on at time $t=0$. The former is the ``canonical\textquotedblright{}
signal used by Sommerfeld and Brillouin. Both cases can be jointly
studied by considering the complex input signal $\tilde{v}(0,t)=u_{H}(t)\exp\left(i\omega_{c}t\right)$.
Its convolution with the Gaussian impulse response gives for the corresponding
output signal 
\begin{equation}
\tilde{v}(\ell,t)\thickapprox\frac{1}{2}\left[1+\mathrm{erf}\left(\beta t+iy\right)\right]e^{-y^{2}}e^{i\omega_{c}t}\label{eq:douze}
\end{equation}
Here $\mathrm{erf}(z)$ indicates the error function, $t$ is the
time retarded by $t_{B}$, and $y=\omega_{c}/\left(2\beta\right)=\sqrt{\alpha_{c}\ell}$,
where $\alpha_{c}$ is a short-hand notation for $\alpha\left(\omega_{c}\right)$.
Taking $t_{B}$ ($1/\beta$) as time origin (time scale), Eq.(\ref{eq:douze})
shows that the output signal only depends on $y$. For sufficiently large $t$, this signal becomes 
\begin{equation}
\tilde{v}_{m}(\ell,t)\thickapprox\frac{1}{2}\left[1+\mathrm{erf}\left(\beta t\right)\right]e^{-\alpha_{c}\ell} e^{i\omega_{c}t}\label{eq:treize}
\end{equation}
and tends to $\tilde{v}_{m}(\ell,t)\thickapprox e^{-\alpha_{c}\ell} e^{i\omega_{c}t}$
when $t\rightarrow\infty$. This part of $\tilde{v}(\ell,t)$ given
by Eq.(\ref{eq:treize}) may naturally be identified to the main signal.
The Brillouin precursor is then given by the remaining part, that
is 
\begin{equation}
\tilde{v}_{p}(\ell,t)\thickapprox\frac{1}{2}\left[\mathrm{erf}\left(\beta t+i\sqrt{\alpha_{c}\ell}\right)\mathrm{-erf}\left(\beta t\right)\right] e^{-\alpha_{c}\ell} e^{i\omega_{c}t}\label{eq:quatorze}
\end{equation}
Since $\omega_{c}=2\beta\sqrt{\alpha_{c}\ell}$, all the previous
signals appear as universal functions of $\beta t$ or of $\omega_{c}t$
that only depend on the optical thickness $\alpha_{c}\ell$ of the
medium at the frequency $\omega_{c}$ of the carrier. To be definite
we consider in the following deionized water and a carrier frequency
$\omega_{c}=2\pi\times10^{9}\,\mathrm{s^{-1}}$, i.e., a period $T_{c}=1\,\mathrm{ns}$
as often considered in the literature \cite{al89,sto01,ou05,ou09}.
We, however, emphasize that our analytical results are quite general.
\begin{figure}[h]
\begin{centering}
\includegraphics[width=85mm]{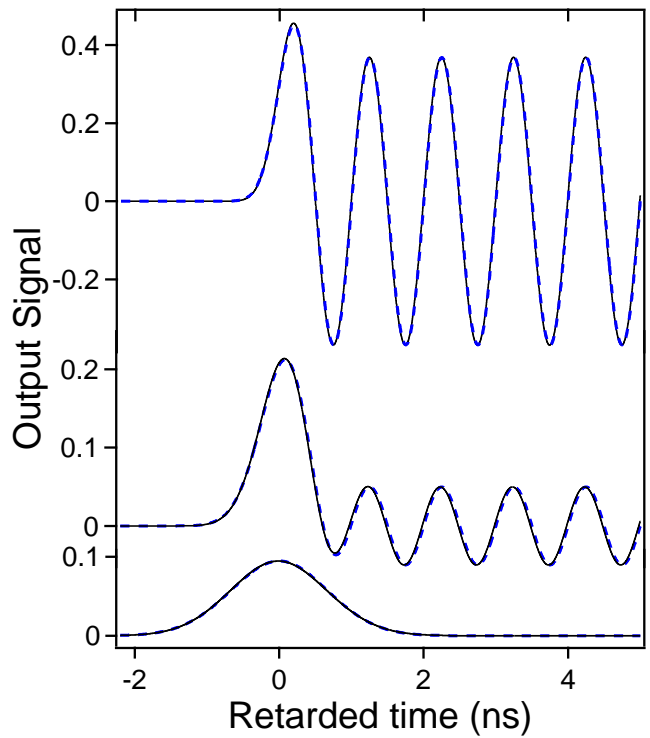} 
\par\end{centering}

\protect\caption{(Color online) Responses to the input signal $u_{H}(t)\sin(\omega_{c}t)$
as a function of the time retarded by $t_{B}$, for $\alpha_{c}\ell=1,\,3\:\mathrm{and}\:10$
(from top to bottom). The solid (dashed) lines give the exact numerical
solution (our analytical solution in terms of error functions). Note
that for the carrier frequency considered ($\omega_{c}/(2\pi)=1\:\mathrm{\mathrm{n}s^{-1}}$),
the optical thickness $\alpha_{c}\ell=1$ is attained for $\ell=21.6\:\mathrm{cm}$.\label{fig:SignalAnalyticVsFFTSin}}
\end{figure}

\begin{figure}[h]
\begin{centering}
\includegraphics[width=85mm]{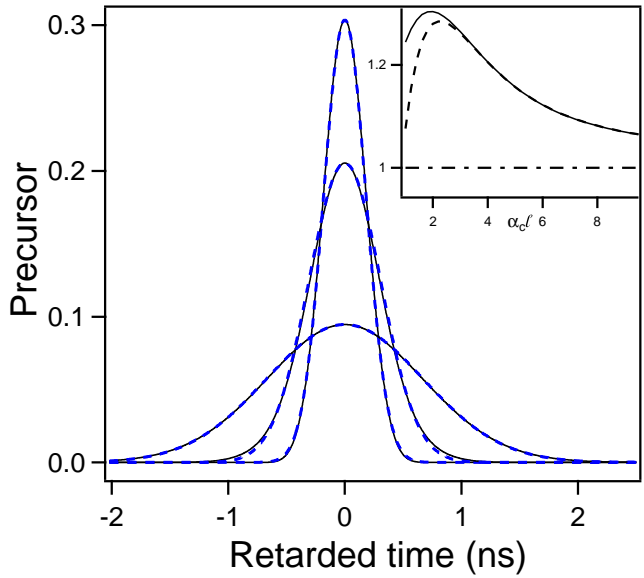} 
\par\end{centering}
\protect\caption{(Color online) Analytical forms of the precursor contribution to the
signals of Fig. \ref{fig:SignalAnalyticVsFFTSin} (solid lines) and
their approximation by the Gaussian given in Eq. (\ref{eq:dixhuit}) (dashed
lines). Inset : ratios of $\beta_{p}$ (solid line) and $A_{p}$ (dashed
line) over their asymptotic values as a function of the optical thickness.\label{fig:PrecursorAnalyticVsApprSin}}
\end{figure}

The responses to the canonical input signal $u_{H}(t)\sin\left(\omega_{c}t\right)$
are obtained by taking the imaginary part of Eqs. (\ref{eq:douze}à)-(\ref{eq:quatorze}).
Figure \ref{fig:SignalAnalyticVsFFTSin} shows the output signals
$v(\ell,t)$ obtained for different values of $\alpha_{c}\ell$. As
expected, they reproduce very well the exact signals obtained by a
FFT calculation of the inverse Fourier transform of $H(\ell,\omega)V(0,\omega)$.
For $\alpha_{c}\ell=1$ ($\ell\approx21.6\,\mathrm{cm}$), the precursor
appears as a small overshot on $v(\ell,t)$, which is hardly visible
for smaller optical thicknesses. For $\alpha_{c}\ell=3$, we have
an example of well developed precursor dominating the main signal.
Finally for $\alpha_{c}\ell=10$, the main signal is extremely small
and only the precursor remains visible. Figure \ref{fig:PrecursorAnalyticVsApprSin}
shows the corresponding precursors alone, as defined by Eq. (\ref{eq:quatorze}).
Owing to the symmetry properties of the real and imaginary parts of
the error function, they are even functions of $t$. Their peak amplitude
is simply 
\begin{equation}
A_{p}(\ell)=\mathrm{Im}\left[\tilde{v}_{p}(\ell,0)\right]=\frac{\mathrm{erf}\left(i\sqrt{\alpha_{c}\ell}\right)}{2i} e^{-\alpha_{c}\ell},\label{eq:seize}
\end{equation}
whereas their area $S_{p}\left(\ell\right)$ , also deduced from Eq. (\ref{eq:quatorze}),
reads as 
\begin{equation}
S_{p}\left(\ell\right)=\frac{\left(1-e^{-2\alpha_{c}\ell}\right)}{\omega_{c}}.\label{eq:dixsept}
\end{equation}
Remarkably enough, their shape is practically Gaussian. The precursor
part of the output signal can thus be written under the approximate form 
\begin{equation}
v_{p}\left(\ell,t\right)=A_{p}(\ell)\,e^{-\beta_{p}^{2}t^{2}},\label{eq:dixhuit}
\end{equation}
where the parameter $\beta_{p}$ is easily determined by combining
Eqs. (\ref{eq:seize}) and (\ref{eq:dixsept}). We get 
\begin{equation}
\beta_{p}=\frac{\sqrt{\pi\alpha_{c}\ell}\mathrm{\,erf}\left(i\sqrt{\alpha_{c}\ell}\right)}{i\left(e^{\alpha_{c}\ell}-e^{-\alpha_{c}\ell}\right)}\beta.\label{eq:dixneuf}
\end{equation}
As shown in Fig. \ref{fig:PrecursorAnalyticVsApprSin}, Eq. (\ref{eq:dixhuit}),
with $A_{p}$ and $\beta_{p}$ given by Eq. (\ref{eq:seize}) and (\ref{eq:dixneuf}),
provides an excellent approximation of the exact result. It strictly
gives the exact result when $\alpha_{c}\ell\rightarrow\infty$. From
the asymptotic form of the error function \cite{ni10}, we then get
$A_{p}(\ell)=\frac{1}{2\sqrt{\pi\alpha_{c}\ell}}=\frac{\beta}{\omega_{c}\sqrt{\pi}}$
and $\beta_{p}=\beta$, that is, 
\begin{equation}
v_{p}\left(\ell,t\right)=\frac{\beta}{\omega_{c}\sqrt{\pi}} e^{-\beta^{2}t^{2}}=\frac{e^{-\beta^{2}t^{2}}}{2\sqrt{\pi\alpha_{c}\ell}}.\label{eq:vingt}
\end{equation}
The well-known law $A_{p}(\ell)\propto1/\sqrt{\ell}$ is retrieved
but we remark that it only holds for very large optical thicknesses.
When $\alpha_{c}\ell$ is only large (such that $e^{-\alpha_{c}\ell}\ll1$),
it is possible to obtain a better approximation of $v_{p}\left(\ell,t\right)$
by considering one more term in the asymptotic expansion of the error
function \cite{ni10}. This leads to a simple multiplication of $\beta$
by $\sqrt{1+1/\left(\alpha_{c}\ell\right)}$ in the first form of
Eq. (\ref{eq:vingt}). The inset of Fig. \ref{fig:PrecursorAnalyticVsApprSin}
more generally shows how the ratios of $\beta_{p}$ and $A_{p}$ over their asymptotic limit (respectively
$\beta$ and $\frac{\beta}{\omega_{c}\sqrt{\pi}}$) vary as a function of $\alpha_{c}\ell$.

We examine briefly the case where $v(0,t)=u_{H}(t)\cos\left(\omega_{c}t\right)$.
Such an input field originates a class of precursors whose amplitude
depends on $\ell$ according to a law differing from the previous
one. We incidentally remark that, for $\omega_{c}=0$ (no carrier),
the output signal is nothing other than the step response of the medium
\cite{pap87} that reads as $a(\ell,t)=\left[1+erf\left(\beta t\right)\right]/2$,
a result in agreement with the observations reported in Refs \cite{fa97,sto01}.
For $\omega_{c}\neq0$ , the output signals are obtained by taking
the real parts of Eqs. (\ref{eq:douze})-(\ref{eq:quatorze}). 
\begin{figure}[h]
\begin{centering}
\includegraphics[width=85mm]{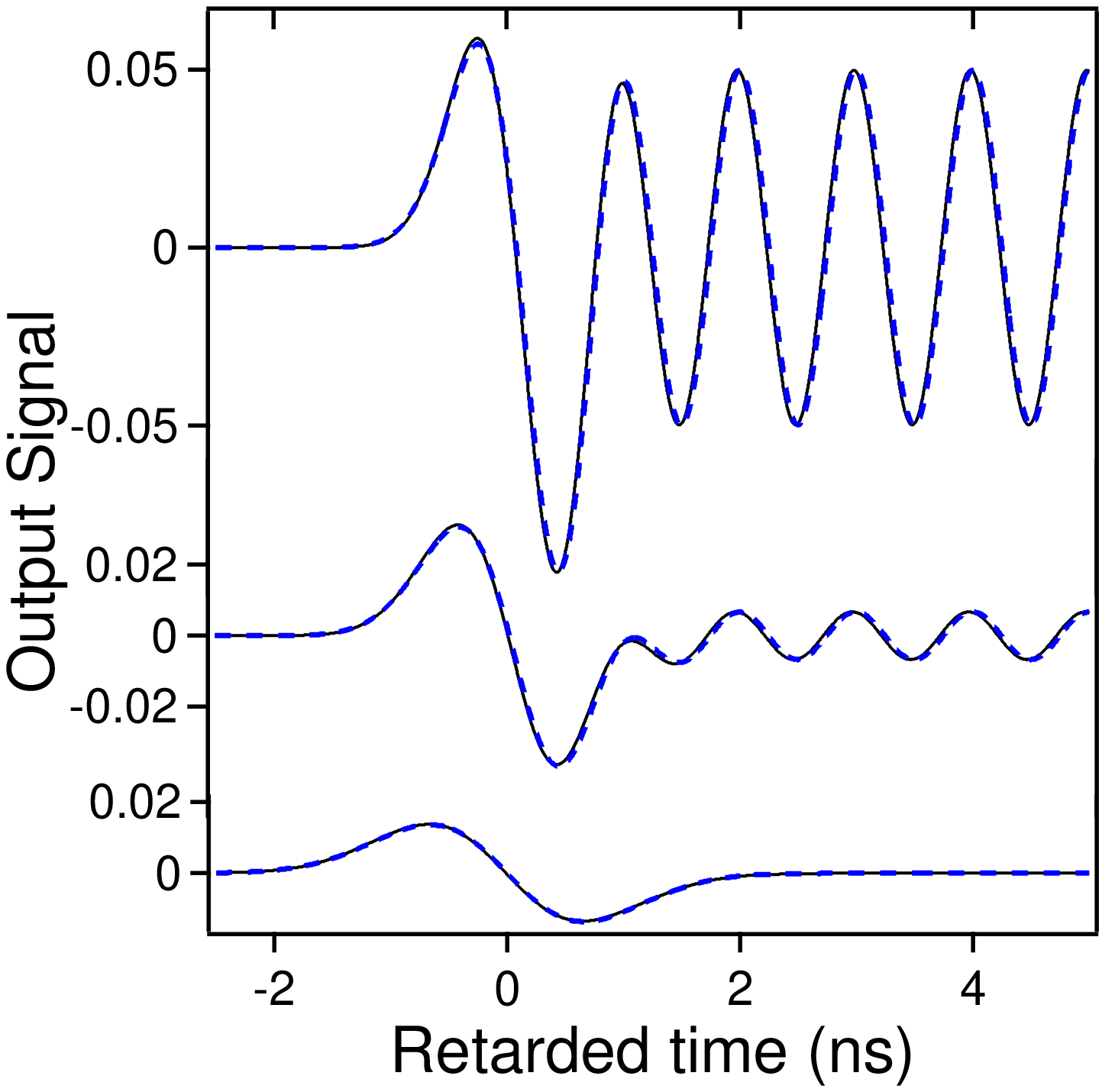} 
\par\end{centering}
\protect\caption{(Color online) Same as Fig. \ref{fig:SignalAnalyticVsFFTSin} when
$v(0,t)=u_{H}(t)\cos(\omega_{c}t)$ for $\alpha_{c}\ell=3,\,5,\:\mathrm{and}\:10$
(from top to bottom).\label{fig:SignalAnalyticVsFFTCos}}
\end{figure}
Figure \ref{fig:SignalAnalyticVsFFTCos} shows the output signals
$v(\ell,t)$ obtained for $\alpha_{c}\ell=3$, $5$, and $10$. The
precursor appears for propagation distances larger than in the previous
case and is not discernible from the main signal when $\alpha_{c}\ell=1$.
The output signals $v(\ell,t)$ derived from Eq. (\ref{eq:douze})
again fit very well the exact signals obtained by FFT. The contribution
$v_{p}(\ell,t)$ of the precursor to the output signal (Fig. \ref{fig:PrecusorAnalyticVsApprCos})
is now an odd function of $t$. Its slope at $t=0$ can be determined
by expanding Eq. (\ref{eq:quatorze}) at the first order in $t$. We
get $\dot{v}_{p}\left(\ell,0\right)=-B_{p}$, with 
\begin{equation}
B_{p}=\beta\left[\sqrt{\alpha_{c}\ell}\frac{\mathrm{erf}\left(i\sqrt{\alpha_{c}\ell}\right)}{i}e^{-\alpha_{c}\ell}-\frac{1-e^{-\alpha_{c}\ell}}{\sqrt{\pi}}\right]>0.\label{eq:vingtdeux}
\end{equation}
\begin{figure}[h]
\begin{centering}
\includegraphics[width=85mm]{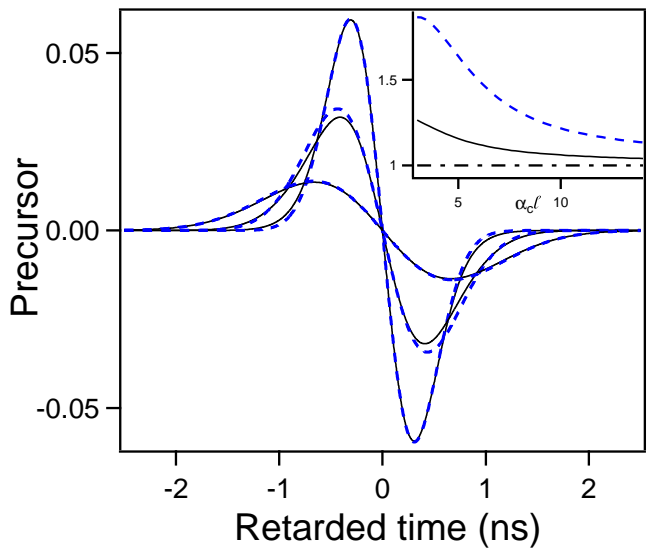} 
\par\end{centering}
\protect\caption{(Color online) Analytical forms of the precursor contribution to the
signals of Fig. \ref{fig:SignalAnalyticVsFFTCos} (solid lines) and
their approximation by the Gaussian derivative given Eq. (\ref{eq:vingttrois})
(dashed lines). Inset is the same that of Fig. \ref{fig:PrecursorAnalyticVsApprSin}.\label{fig:PrecusorAnalyticVsApprCos}}
\end{figure}
The precursors are now well fitted by derivatives of Gaussian (Fig. \ref{fig:PrecusorAnalyticVsApprCos})
and can be written as $v_{p}(\ell,t)\thickapprox-B_{p}t e^{-\gamma_{p}^{2}t^{2}}$.
In fact, $\gamma_{p}$, achieving the best fit, never considerably
differs from $\beta_{p}$. The precursor takes then the approximate
form 
\begin{equation}
v_{p}(\ell,t)\approx -B_{p}t e^{-\beta_{p}^{2}t^{2}}=\frac{B_{p}}{2\beta_{p}^{2}}\frac{d}{dt}e^{-\beta_{p}^{2}t^{2}}\label{eq:vingttrois}
\end{equation}
with a peak amplitude 
\begin{equation}
A_{p}(\ell)=\frac{B_{p}}{\beta_{p}\sqrt{2e}} .\label{eq:vingtquatre}
\end{equation}
As shown in Fig. \ref{fig:PrecusorAnalyticVsApprCos}, Eq. (\ref{eq:vingttrois})
provides a satisfactory approximation of the precursor. It gives the
exact result when $\alpha_{c}\ell\rightarrow\infty$. In this limit,
$\beta_{p}=\beta$, $B_{p}=2\beta^{3}/\left(\omega_{c}^{2}\sqrt{\pi}\right)$,
$A_{p}(\ell)=1/\left(2\alpha_{c}\ell\sqrt{2e\pi}\right)$ and 
\begin{multline}
v_{p}(\ell,t)=\left(\frac{\beta}{\omega_{c}^{2}\sqrt{\pi}}\right)\frac{d}{dt}\left(e^{-\beta^{2}t^{2}}\right)=\\
-\left(\frac{1}{2\alpha_{c}\ell\sqrt{\pi}}\right)\beta t\,e^{-\beta^{2}t^{2}}.\label{eq:vingtcinq}
\end{multline}
Since $\omega_{c}u_{H}(t)\cos\left(\omega_{c}t\right)=\frac{d}{dt}\left[u_{H}(t)\sin\left(\omega_{c}t\right)\right]$
and $v_{p}(\ell,t)=v(\ell,t)$ when $\alpha_{c}\ell\rightarrow\infty$,
this precursor is the time derivative of that obtained in the canonical
case {[}see Eq.(\ref{eq:vingt}){]} divided by $\omega_{c}$. Its
amplitude now scales as $1/\ell$ instead as $1/\sqrt{\ell}$ in the
previous case. As shown in the inset of Fig. \ref{fig:PrecusorAnalyticVsApprCos},
this law only holds for very large optical thicknesses.

From the results obtained with the input signals $u_{H}(t)\sin\left(\omega_{c}t\right)$
and $u_{H}(t)\cos\left(\omega_{c}t\right)$, the responses to the
more general input signal $u_{H}(t)\sin\left(\omega_{c}t+\varphi\right)$
are easily determined. The precursor is then a superposition of a
Gaussian and a Gaussian derivative with relative amplitudes depending
on $\varphi$ and on the propagation distance.

\begin{figure}[h]
\begin{centering}
\includegraphics[width=85mm]{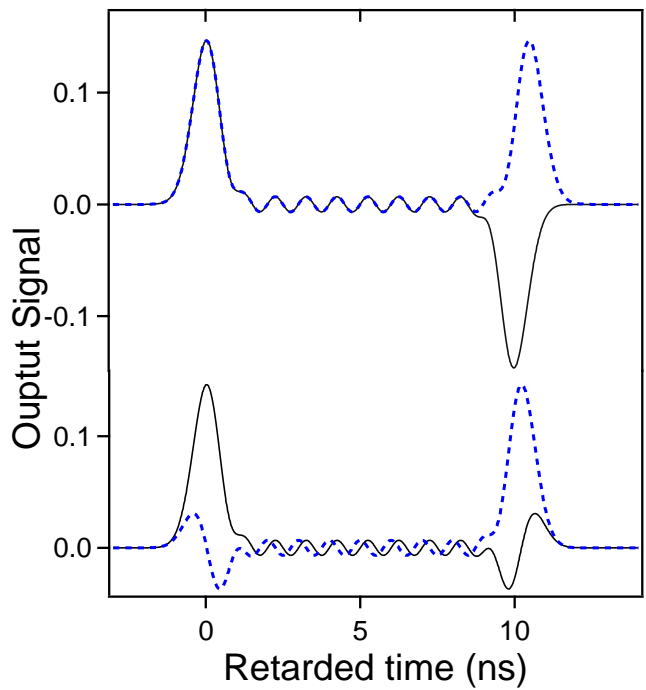} 
\par\end{centering}
\protect\caption{(Color online) Responses to input signals with a long rectangular
envelope for $\alpha_{c}\ell=5$. The upper curves are obtained for
$v(0,t)\propto\sin(\omega_{c}t)$ with $T=10\,T_{c}$ (solid line)
and $T=10.5\,T_{c}$ (dashed line). The lower curves are obtained
for $T=10.25\,T_{c}$ with $v(0,t)\propto\sin(\omega_{c}t)$ (solid
line) and $v(0,t)\propto\cos(\omega_{c}t)$ (dotted line). Note that,
in the latter case, the postcursor can completely dominate the precursor.\label{fig:PrecuseurRectangularSinePulse}}
\end{figure}

In real or numerical experiments, the input signal has obviously a
finite duration. One generally considers input signals with a rectangular
envelope of duration $T$ long enough to avoid having the precursors
generated by the rise and the fall of the pulse overlap. The corresponding
complex signal reads as $\tilde{v}_{R}(0,t)=\left[u_{H}(t)-u_{H}(t-T)\right]e^{i\omega_{c}t}$,
that is 
\begin{equation}
\tilde{v}_{R}(0,t)=u_{H}(t)\,e^{i\omega_{c}t}-e^{i\omega_{c}T}u_{H}(t-T)\,e^{i\omega_{c}\left(t-T\right)}.\label{eq:vingtsix}
\end{equation}
It generates the output signal 
\begin{equation}
\tilde{v}_{R}(\ell,t)=\tilde{v}(\ell,t)-e^{i\omega_{c}T}\,\tilde{v}(\ell,t-T),\label{eq:vingtsept}
\end{equation}
where $\tilde{v}(\ell,t)$ is given by Eq. (\ref{eq:douze}). No matter
$T$, the mains signals generated by the rise and the fall of the
pulse destructively interferes when $\beta\left(t-T\right)\gg\sqrt{\alpha_{c}\ell}$.
On the other hand, the second precursor (a postcursor since it follows
the main signal) generally differs from the first one owing to the
phase factor $e^{i\omega_{c}T}$. Precursor and postcursor
are identical only when $T=\left(2n+1\right)T_{c}/2$, where $n$ is
an integer. They have the same shape and amplitude but opposite signs
when $T=nT_{c}$. Both cases have been evidenced in the experiments
reported in Ref \cite{sto01}. The largest difference between precursor
and postcursor is attained when $T=\left(2n+1\right)T_{c}/4$. In
this case $e^{i\omega_{c}T}=\left(-1\right)^{n}i$. For an
input signal $v(0,t)=\left[u_{H}(t)-u_{H}(t-T)\right]\,\sin\left(\omega_{c}t\right)$,
we get an output signal 
\begin{equation}
v_{R}(\ell,t)=\mathrm{Im}\left[\tilde{v}(\ell,t)\right]-\left(-1\right)^{n}\mathrm{Re}\left[\tilde{v}(\ell,t-T)\right]\label{eq:vingthuit}
\end{equation}
The precursor is then Gaussian {[}see Eq. (\ref{eq:dixhuit}){]} whereas
the postcursor is a Gaussian derivative {[}see Eq. (\ref{eq:vingttrois}){]}
of smaller amplitude. The opposite occurs for an input signal $v(0,t)=\left[u_{H}(t)-u_{H}(t-T)\right]\,\cos\left(\omega_{c}t\right)$.
These various behaviors are illustrated in Fig. \ref{fig:PrecuseurRectangularSinePulse}. 

\begin{figure}[h]
\begin{centering}
\includegraphics[width=85mm]{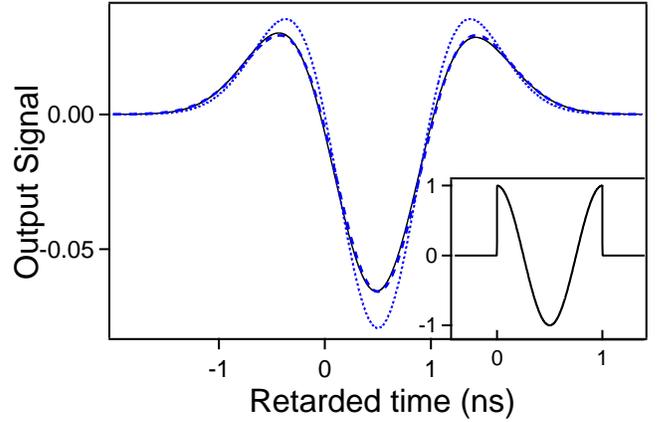} 
\par\end{centering}
\protect\caption{(Color online) Response to the input signal $\left[u_{H}(t)-u_{H}(t-T_{c})\right]\cos(\omega_{c}t)$
for $\alpha_{c}\ell=5$. The solid, dashed and dotted lines are the
exact FFT solution, the approximate analytical solution given Eq. (\ref{eq:trentetrois}),
and the asymptotic solution given Eq. (\ref{eq:trente}), respectively.
Inset: input signal.\label{fig:OneperiodeRectangularCosPulse}}
\end{figure}

Equation (\ref{eq:vingtsept}) remains valid when the two precursors
overlap and interfere to give a unique transient that may be considered
as a generalized precursor. A remarkable behavior is obtained for
an input signal $\left[u_{H}(t)-u_{H}(t-T_{c})\right]\,\cos\left(\omega_{c}t\right)$.
A good overlapping of the precursors is achieved when $1/\beta\gg T_{c}$,
that is in the asymptotic limit. We then get 
\begin{equation}
v(\ell,t)\thickapprox v_{p}(\ell,t)-v_{p}(\ell,t-T_{c})\approx T_{c}\dot{v}_{p}\left(\ell,t-\frac{T_{c}}{2}\right),\label{eq:vingtneuf}
\end{equation}
where $v_{p}(\ell,t)$ is given by Eq. (\ref{eq:vingtcinq}), and finally
\begin{multline}
v(\ell,\theta)=\left(\frac{2\beta\sqrt{\pi}}{\omega_{c}^{3}}\right)\frac{d^{2}}{d\theta^{2}}\left(e^{-\beta^{2}\theta^{2}}\right)=\\
-A_{\infty}\left(1-2\beta^{2}\theta^{2)}\right)e^{-\beta^{2}\theta^{2}},\label{eq:trente}
\end{multline}
where $\theta=t-T_{c}/2$ and $A_{\infty}=\left(\sqrt{\pi}/2\right)\left(\alpha_{c}\ell\right)^{-3/2}$.
The output signal has now the shape of a Gaussian second-derivative
centered at $t=T_{c}/2$ and a peak amplitude $A_{\infty}$ scaling
as $\ell^{-3/2}$. For any optical thickness, it is easily shown from
Eq.(\ref{eq:vingtsept}) that $v(\ell,\theta)$ is an even function
of $\theta$ with a minimum at $\theta=0$ of algebraic amplitude
$-A$, with 
\begin{equation}
A=e^{-\alpha_{c}\ell}\,\mathrm{Re}\left[\mathrm{erf}\left(\frac{\pi\beta}{\omega_{c}}+i\frac{\omega_{c}}{2\beta}\right)\right]\label{eq:trenteetun}
\end{equation}
that tends to $A_{\infty}$ when $\alpha_{c}\ell\rightarrow\infty$.
It also results from Eq. (\ref{eq:vingtsept}) that 
\begin{equation}
v\left(\ell,\theta\right)=\frac{e^{-\alpha_{c}\ell}}{2}\,\mathrm{Re}\left[\mathrm{erf}\left(\frac{2\pi\beta}{\omega_{c}}+i\frac{\omega_{c}}{2\beta}\right)\right]\label{eq:trentedeux}
\end{equation}
for $\theta=\pm\pi/\omega_{c}$, that is, at the switching times of
the input signal delayed by $t_{B}$. Unexpectedly enough, extensive
numerical simulations show that $v\left(\ell,\theta\right)$ is very
well fitted by a Gaussian second derivative as soon as $\alpha_{c}\ell>1$.
The output signal then reads as 
\begin{equation}
v\left(\ell,\theta\right)=-A\left(1-2\eta^{2}\beta^{2}\theta^{2}\right)e^{-\eta^{2}\beta^{2}\mathit{\theta}^{2}},\label{eq:trentetrois}
\end{equation}
where $A$ is given by Eqs. (\ref{eq:trenteetun}) and $\eta$ is a
parameter obtained by combing Eq. (\ref{eq:trenteetun}) and (\ref{eq:trentedeux}).
Figure \ref{fig:OneperiodeRectangularCosPulse} shows the result obtained
by this method for $\alpha_{c}\ell=5$. It perfectly fits the exact
numerical result obtained by FFT. The asymptotic result of Eq. (\ref{eq:trente})
is also given for comparison. In spite of the moderate value of the
optical thickness, it is reasonable approximation of the exact
result. 

\section{RESPONSE TO SINE-WAVES WITH LINEARLY VARYING AMPLITUDE \label{sec:Trapeze} }

We consider in this section input signals of the form $\left(t/T_{r}\right)u_{H}(t)\sin\left(\omega_{c}t\right)$
or $\left(t/T_{r}\right)u_{H}(t)\cos\left(\omega_{c}t\right)$ where
$T_{r}$ is the time for which their amplitude equals 1 (rise time).
Such signals can generate precursors preceding significantly the rise
of the main signal. The corresponding complex signal reads as $\tilde{v}(0,t)=\left(t/T_{r}\right)u_{H}(t)\,e^{i\omega_{c}t}$.
By convoluting this signal with the Gaussian impulse response, we get 
\begin{multline}
\tilde{v}(\ell,t)=\frac{e^{-\beta^{2}t^{2}}}{2\beta T_{r}\sqrt{\pi}}+\frac{e^{-\alpha_{c}\ell}}{2\beta T_{r}}\times\\
\left\{ \left(\beta t+i\sqrt{\alpha_{c}\ell}\right)\left[1+\mathrm{erf}\left(\beta t+i\sqrt{\alpha_{c}\ell}\right)\right]\right\} e^{i\omega_{c}t}.\label{eq:trentequatre}
\end{multline}
Proceeding as in the previous section, we get for the main signal
and the precursor 
\begin{equation}
\tilde{v}_{m}(\ell,t)=\frac{e^{-\alpha_{c}\ell}}{2\beta T_{r}}\left\{ \left(\beta t+i\sqrt{\alpha_{c}\ell}\right)\left[1+\mathrm{erf}\left(\beta t\right)\right]\right\} e^{i\omega_{c}t},\label{eq:trentecinq}
\end{equation}
\begin{equation}
\tilde{v}_{p}(\ell,t)=\tilde{v}(\ell,t)-\tilde{v}_{m}(\ell,t).\label{eq:trentesix}
\end{equation}
All these signals are inversely proportional to $T_{r}$, and $\omega_{c}$
being fixed, it is convenient to normalize them to $1/\left(\omega_{c}T_{r}\right)$.
\begin{figure}[h]
\begin{centering}
\includegraphics[width=85mm]{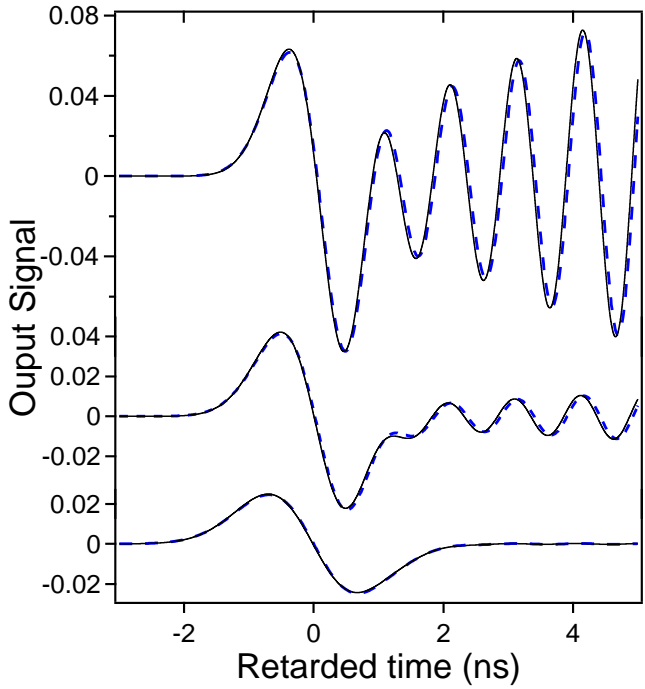} 
\par\end{centering}
\protect\caption{(Color online) Responses to the input signal $(t/T_{r})u_{H}(t)\sin(\omega_{c}t)$
normalized to $1/(\omega_{c}T_{r})$ for $\alpha_{c}\ell=6$, $8$, and $12$
(from top to bottom). The solid (dashed) line is the FFT exact solution
(the approximate analytical solution in terms of error functions).\label{fig:LineaireSin}}
\end{figure}

\begin{figure}[h]
\begin{centering}
\includegraphics[width=85mm]{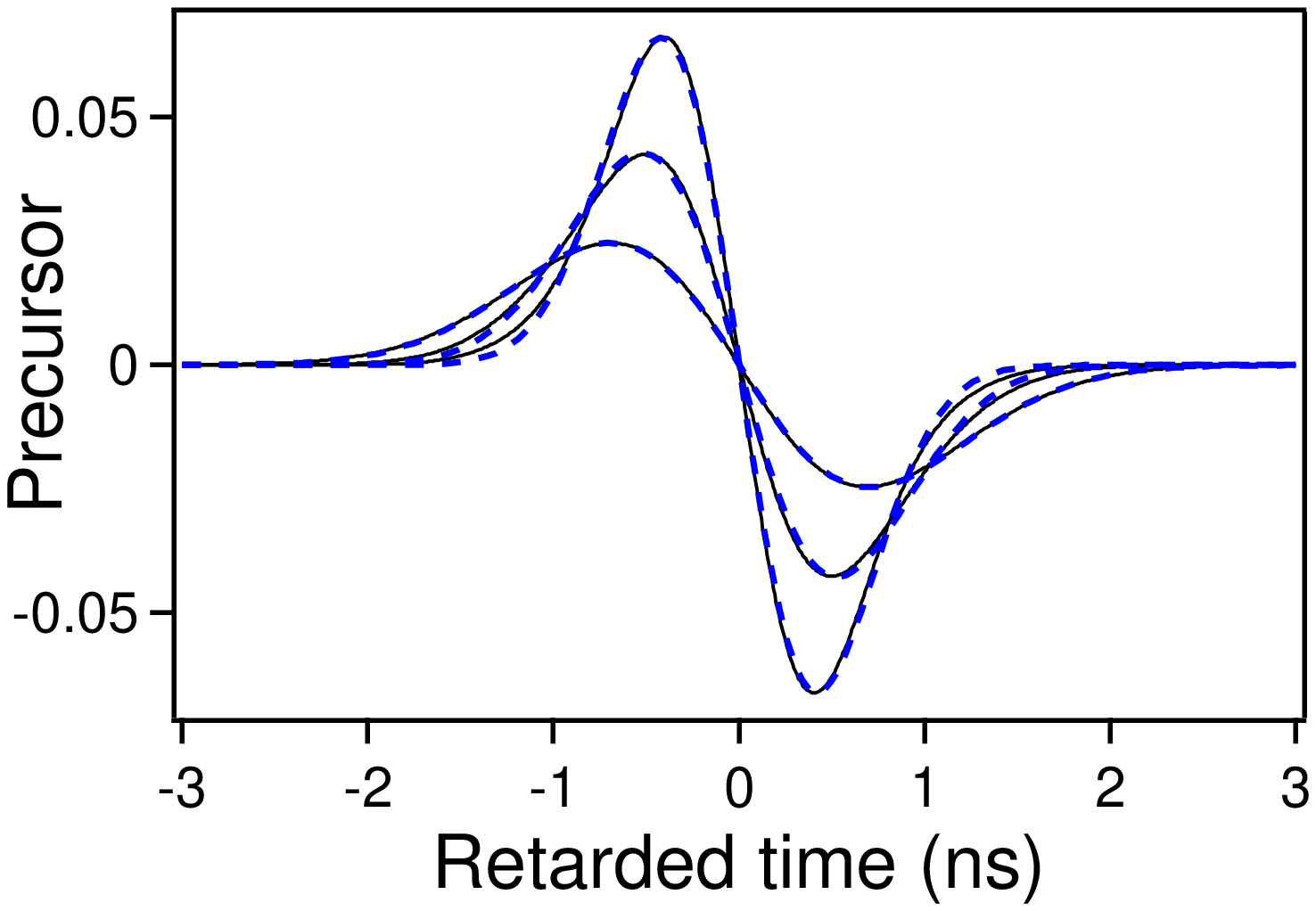} 
\par\end{centering}
\protect\caption{(Color online) Analytical forms of the precursor contribution to the
signals of Fig. \ref{fig:LineaireSin} (solid lines) and their best
fit by Gaussian derivatives (dashed lines).\label{fig:Precusor_Sin-vs-FIT}}
\end{figure}

Figure \ref{fig:LineaireSin} shows the output signal $v(\ell,t)=\mathrm{Im}\left[\tilde{v}(\ell,t)\right]$
generated by $v(0,t)=\left(t/T_{r}\right)u_{H}(t)\sin\left(\omega_{c}t\right)$
for different optical thicknesses. The precursor becomes discernible
from the main field for $\alpha_{c}\ell\geq6$ and is fully separated
from it for $\alpha_{c}\ell=12$. In every case the analytical result
derived from Eq. (\ref{eq:trentequatre}) fits very well the exact
numerical result obtained by FFT. The contribution $v_{p}(\ell,t)=\mathrm{Im}\left[\tilde{v_{p}}(\ell,t)\right]$
of the precursor to the output signal is now an odd function of the
retarded time $t$ and, for $\alpha_{c}\ell\geq6$ , is well fitted
by a Gaussian derivative (Fig. \ref{fig:Precusor_Sin-vs-FIT}). This
result is rigorous when $\alpha_{c}\ell\rightarrow\infty$ . By means
of asymptotic expansions of the error function limited to its first
non zero term in $1/\sqrt{\alpha_{c}\ell}$, we then get 
\begin{multline}
v_{p}(\ell,t)=-\left(\frac{1}{\omega_{c}T_{r}}\right)\left(\frac{1}{\alpha_{c}\ell\sqrt{\pi}}\right)\beta t\,e^{-\beta^{2}t^{2}}=\\
\left(\frac{2\beta}{\omega_{c}^{3}T_{r}\sqrt{\pi}}\right)\frac{d}{dt}\left(e^{-\beta^{2}t^{2}}\right),\label{eq:trentesept}
\end{multline}
the precursor amplitude scaling as $1/\ell$. 

\begin{figure}[h]
\begin{centering}
\includegraphics[width=85mm]{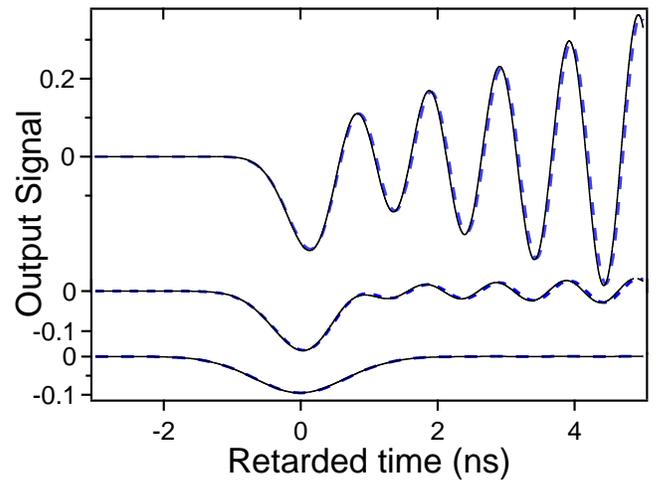} 
\par\end{centering}
\protect\caption{(Color online) Same as Fig. \ref{fig:LineaireSin} with $v(0,t)=(t/T_{r})u_{H}(t)\cos(\omega_{c}t)$
for $\alpha_{c}\ell=4.5$, $7$, and $12$ (from top to bottom).\label{fig:LinaireCos}}
\end{figure}

\begin{figure}[h]
\begin{centering}
\includegraphics[width=85mm]{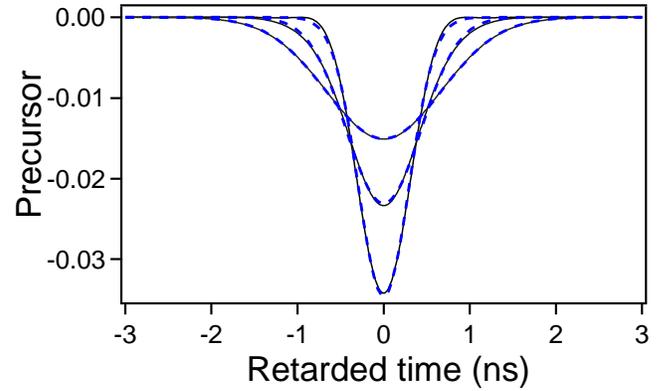} 
\par\end{centering}
\protect\caption{(Color online) Analytical forms of the precursor contribution to the
signals of Fig. \ref{fig:LinaireCos} (solid lines) and their best fit
by Gaussians (dashed lines).\label{fig:Precusor_Cos-vs-FIT}}
\end{figure}

When $v(0,t)=\left(t/T_{r}\right)u_{H}(t)\cos\left(\omega_{c}t\right)$
, the overall output signal and the precursor read as $v(\ell,t)=\mathrm{Re}\left[\tilde{v}(\ell,t)\right]$
and $v_{p}(\ell,t)=\mathrm{Re}\left[\tilde{v_{p}}(\ell,t)\right]$
. The precursor becomes discernible from the main field for $\alpha_{c}\ell\geq4.5$
(Fig. \ref{fig:LinaireCos}) and, again owing to the symmetry properties
of the error function, is an even function of $t$. 
At $t=0$, $v_{p}(\ell,t)$ has the remarkable value 
\begin{equation}
v_{p}(\ell,0)=-A=\frac{1}{\omega_{c}T_{r}}\left[\sqrt{\frac{\alpha_{c}\ell}{\pi}}-\alpha_{c}\ell\frac{\mathrm{erf}\left(i\sqrt{\alpha_{c}\ell}\right)}{i}e^{-\alpha_{c}\ell}\right],\label{eq:trentehuit}
\end{equation}
where $A>0$ as soon as $\alpha_{c}\ell\geq1$. For $\alpha_{c}\ell\geq4.5$,
the precursor is well fitted by a (negative) Gaussian (Fig. \ref{fig:Precusor_Cos-vs-FIT}).
This solution is exact when $\alpha_{c}\ell\rightarrow\infty$ and
we then get from Eq. (\ref{eq:trentesix}) 
\begin{equation}
v_{p}(\ell,t)=-\frac{\beta}{\omega_{c}^{2}T_{r}\sqrt{\pi}}e^{-\beta^{2}t^{2}}=-A_{\infty}e^{-\beta^{2}t^{2}},\label{eq:trenteneuf}
\end{equation}
where $A_{\infty}=1/\left(2\omega_{c}T_{r}\sqrt{\pi\alpha_{c}\ell}\right)$
is the limit of $A$ when $\alpha_{c}\ell\rightarrow\infty$ and scales
as $1/\sqrt{\ell}$.

When the precursors are generated by a single discontinuity of the
input signal, their shape in the asymptotic limit strongly depends
on the order of this discontinuity. Equations (\ref{eq:vingt}), (\ref{eq:vingtcinq}),
(\ref{eq:trentesept}), and (\ref{eq:trenteneuf}) lead us to conjecture
that the precursor has a Gaussian (Gaussian derivative) shape when
the discontinuity order is odd (even). Complementary calculations
made when $v(0,t)=\left(t/T_{r}\right)^{n}u_{H}(t)\sin\left(\omega_{c}t\right)$
or $\left(t/T_{r}\right)^{n}u_{H}(t)\cos\left(\omega_{c}t\right)$
for $n=2$ and $n=3$ support this conjecture but it should be remarked
that the precursor amplitude is a rapidly decreasing function of $n$
.

The effect of the rise time of the input signal on the precursors
is generally studied by considering input signals whose amplitude
linearly increases during a time $T_{r}\leq T_{c}$ and is then maintained
constant \cite{al89,sto01,ou09,ou12}. The corresponding complex signal
may be written as 
\begin{multline}
\tilde{v}_{L}(0,t)=\frac{t}{T_{r}}u_{H}(t)\,e^{i\omega_{c}t}\\
-e^{i\omega_{c}T_{r}}\frac{t-T_{r}}{T_{r}}u_{H}(t-T_{r})\,e^{i\omega_{c}(t-T_{r})}.\label{eq:quarante}
\end{multline}
It generates the output signal 
\begin{equation}
\tilde{v}_{L}(\ell,t)=\tilde{v}(\ell,t)-e^{i\omega_{c}T_{r}}\tilde{\,v}(\ell,t-T_{r}),\label{eq:quaranteetun}
\end{equation}
where $\tilde{v}(\ell,t)$ is given by Eq. (\ref{eq:trentequatre}).
For $\beta\left(t-T_{r}\right)\gg\alpha_{c}\ell$, Eq. (\ref{eq:quaranteetun})
is reduced to $\tilde{v}_{L}(\ell,t)=e^{-\alpha_{c}\ell}e^{i\omega_{c}t}$,
as expected since the long-term behavior of the main signal should
not depend on the rise time of the input signal. Remarkable behaviors
are obtained in the asymptotic limit when $T_{r}=T_{c}$ and $T_{r}=T_{c}/2$.
When $T_{r}=T_{c}$, the precursors generated by the first and second
discontinuities of the slope of the input signal interfere nearly
destructively and the output signal are again given by Eq. (\ref{eq:vingtneuf})
where $\tilde{v}_{p}(\ell,t)$ is now given by Eq. (\ref{eq:trentesept})
for $v(0,t)\propto\sin\left(\omega_{c}t\right)$ and by Eq. (\ref{eq:trenteneuf})
for $v(0,t)\propto\cos\left(\omega_{c}t\right)$ . The output signals
then reads as 
\begin{multline}
v(\ell,\theta)=\frac{2\beta}{\omega_{c}^{3}\sqrt{\pi}}\frac{d^{2}}{d\theta^{2}}\left(e^{-\beta^{2}\theta^{2}}\right)\\
=-\frac{\left(1-2\beta^{2}\theta^{2}\right)e^{-\beta^{2}\theta^{2}}}{2\left(\alpha_{c}\ell\right)^{3/2}\sqrt{\pi}}\label{eq:quarantedeux}
\end{multline}
(amplitude $\propto\ell^{-3/2}$) in the former case and as 
\begin{equation}
v(\ell,\theta)=-\frac{\beta}{\omega_{c}^{2}\sqrt{\pi}}\frac{d}{d\theta}\left(e^{-\beta^{2}\theta^{2}}\right)=\frac{\beta\theta\,e^{-\beta^{2}\theta^{2}}}{2\alpha_{c}\ell\sqrt{\pi}}\label{eq:quarantetrois}
\end{equation}
(amplitude $\propto1/\ell$) in the latter case. In both cases, $\theta=t-T_{c}/2$
. When $T_{r}=T_{c}/2$, the two precursors interfere constructively.
For $v(0,t)\propto\sin\left(\omega_{c}t\right)$ and $v(0,t)\propto\cos\left(\omega_{c}t\right)$,
we respectively get 
\begin{equation}
v(\ell,\theta)=\frac{4\beta}{\omega_{c}^{2}\pi^{3/2}}\,\frac{d}{d\theta}\left(e^{-\beta^{2}\theta^{2}}\right)=-\frac{2\beta\theta\,e^{-\beta^{2}\mathit{\theta}^{2}}}{\alpha_{c}\ell\pi^{3/2}}\label{eq:quarantequatre}
\end{equation}
(amplitude $\propto1/\ell$) and 
\begin{equation}
v(\ell,\theta)=-\frac{2\beta}{\omega_{c}\pi^{3/2}}\,e^{-\beta^{2}\theta^{2}}=-\frac{e^{-\beta^{2}\theta^{2}}}{\pi^{3/2}\sqrt{\alpha_{c}\ell}}\label{eq:quarantecinq}
\end{equation}
(amplitude $\propto\ell^{-1/2}$), with $\theta=t-T_{c}/4$ in both
cases. 
Figure \ref{fig:LinearPrecusorTr=00003D00003D00003D1=0000260,5Tc}
shows the output signals given by Eqs. (\ref{eq:quaranteetun}) combined
with Eq. (\ref{eq:trentequatre}) in the four cases considered in this
paragraph. The optical thickness $\alpha_{c}\ell=7$ has been chosen
in order that the main signal is dominated by the precursor but remains
visible. The precursors obtained at the asymptotic limit are given
for reference. Though the asymptotic limit is not attained, they provide
a reasonable approximation of the complete signals. Complementary
simulations obviously show that as $\alpha_{c}\ell$ is larger, the fit of the output signals by the asymptotic form of
the precursors is better. Signals as those shown Fig. \ref{fig:LinearPrecusorTr=00003D00003D00003D1=0000260,5Tc}
have been actually observed in the numerical experiments reported
in Refs \cite{al89,sto01,ou09,ou12} but the fact that the precursor shape
may be a Gaussian, a Gaussian first derivative or a Gaussian second derivative
is either not clearly stated or completely overlooked. All these numerical
simulations were made by using a trapezoidal modulation with the same
rise and fall times and a plateau whose duration is much larger than that
of the precursors.
\begin{figure}[h]
\begin{centering}
\includegraphics[width=85mm]{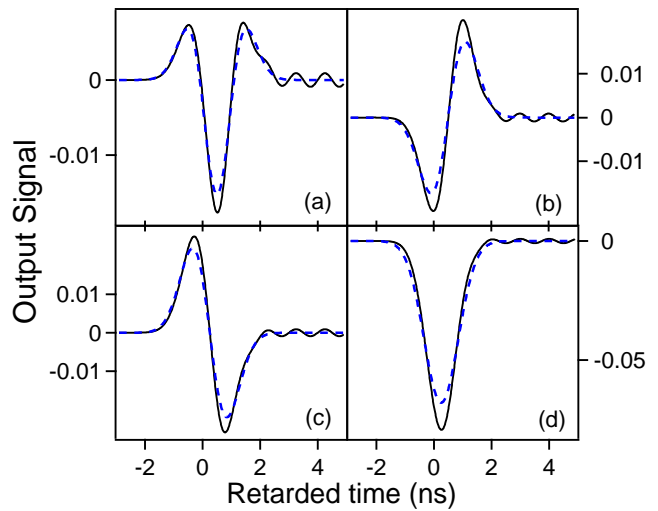} 
\par\end{centering}
\protect\caption{(Color online) Responses (solid line) to input signals with trapezoidal
envelope of rise times $T_{r}=T_{c}$ or $T_{c}/2$ for $\alpha_{c}\ell=7$.
(a) $T_{r}=T_{c}$, $v(0,t)\propto\sin(\omega_{c}t)$; (b) $T_{r}=T_{c}$,
$v(0,t)\propto\cos(\omega_{c}t)$; (c) $T_{r}=T_{c}/2$, $v(0,t)\propto\sin(\omega_{c}t)$;
and (d) $T_{r}=T_{c}/2$, $v(0,t)\propto\cos(\omega_{c}t)$ . The
asymptotic solutions (dashed lines) are given for comparison.\label{fig:LinearPrecusorTr=00003D00003D00003D1=0000260,5Tc}}
\end{figure}
As in the case of a rectangular modulation, depending
on the plateau duration, the precursor and the postcursor may have
the same or opposite signs \cite{al89,sto01,ou09,ou12} and even different
shapes, e.g. a Gaussian associated with a Gaussian first-derivative
or a Gaussian first derivative associated to a Gaussian second derivative.
We incidentally remark that, by using a modulation with rise time,
plateau duration and fall time all equal to $T_{c}$, it would theoretically
be possible to generate a precursor with a Gaussian third derivative
shape and an amplitude scaling as $1/\ell^{2}$. Calculations made
in this case show that the corresponding precursor would emerge from
the main field at distances where it would be too small to be detected
in a real experiment.

\section{GENERAL ASYMPTOTIC PROPERTIES OF THE PRECURSORS\label{sec:AsymptoticProperties}}

The shapes of the precursors obtained at the asymptotic limit and
the dependence of their amplitude with the propagation distance (as
$\ell^{-1/2}$, $\ell^{-1}$ or $\ell^{-3/2}$) are not specific to
the input signals considered in the previous sections. As shown below
they only result from the fact that, in the asymptotic limit, the
width of the transfer function $H(\ell,\omega)$ is infinitely small
compared to that of the Laplace-Fourier transform $V(0,\omega)$ of
the input signal. The latter can then be developed in cumulants by
keeping only the first of them. It reads as $V(0,\omega)=V(0,0)\,\mathrm{e}^{-i\omega t_{0}}=S_{0}e^{-i\omega t_{0}}$
where $S_{0}$ is the area of the input signal and $t_{0}$ its center
of gravity \cite{re1}. We then get $V(\ell,\omega)=S_{0}H(\ell,\omega)e^{-i\omega t_{0}}$
and finally 
\begin{equation}
v(\ell,t)=S_{0}\,h(t-t_{0})\label{eq:quarantecinqbis}
\end{equation}
This result, leading to a precursor amplitude scaling as $\ell^{-1/2}$,
is general and applies in particular to the precursors given by Eqs. (\ref{eq:vingt}), (\ref{eq:trenteneuf}), and (\ref{eq:quarantecinq}), for which the input signals are such that $\left[S_{0},t_{0}\right]=\left[1/\omega_{c},0\right]$,
$\left[-1/\left(\omega_{c}^{2}T_{r}\right),0\right]$ and $\left[-2/\left(\pi\omega_{c}\right),T_{c}/4\right]$,
respectively.

The previous demonstration obviously fails when the input signal has
a zero area. We consider in this case the first antiderivative of
the input signal, namely $^{\left(1\right)}v(0,t)=\int_{-\infty}^{t}v(0,u_{1})\,du_{1}$,
the Fourier transform of which reads as $V(0,\omega)/\left(i\omega\right)$.
Repeating the previous procedure, we get $V(\ell,\omega)=\left(i\omega\right)S_{1}H(\ell,\omega)\mathrm{e}^{-i\omega t_{1}}$
and finally 
\begin{equation}
v(\ell,t)=S_{1}\,\dot{h}(t-t_{1})\label{eq:qurantecinqter}
\end{equation}
where $S_{1}$ and $t_{1}$ are the area and the center of gravity
of $^{\left(1\right)}v(0,t)$ \cite{re1}. This general result, leading
to a precursor amplitude scaling as $1/\ell$, applies in particular
to the precursors given by Eqs. (\ref{eq:vingtcinq}), (\ref{eq:trentesept}),
(\ref{eq:quarantetrois}), and (\ref{eq:quarantequatre}). The
corresponding input signals are actually such that $S_{0}=0$ whereas
$\left[S_{1},t_{1}\right]=\left[1/\omega_{c}^{2},0\right]$, $\left[2/\left(\omega_{c}^{3}T_{r}\right),0\right]$,
$\left[-1/\omega_{c}^{2},T_{c}/2\right]$ and $\left[4/\left(\pi\omega_{c}^{2}\right),T_{c}/4\right]$,
respectively.

\begin{figure}[h]
\begin{centering}
\includegraphics[width=85mm]{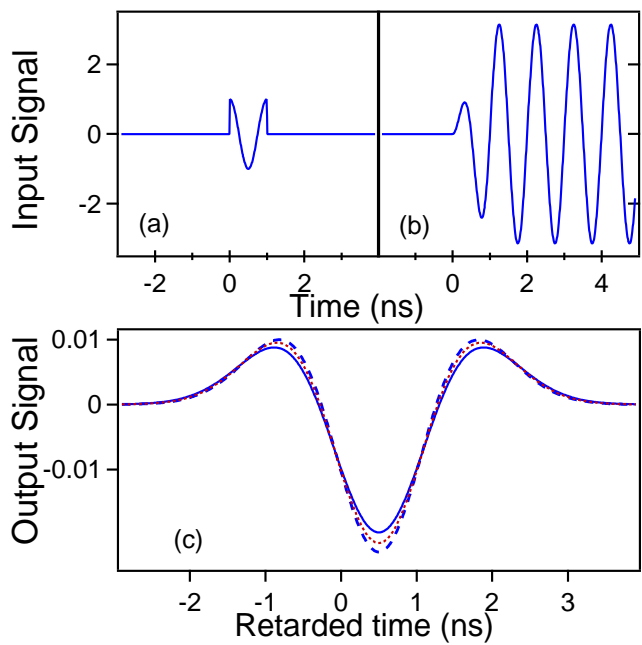} 
\par\end{centering}
\protect\caption{(Color online) Responses to the input signals (a) (solid line) and
(b) (dashed line) for $\alpha_{c}\ell=12$. The asymptotic solution
of Eq. (\ref{eq:quarantecinqquater}) is given for comparison (dotted
line).\label{fig:cestpareil}}
\end{figure}

When $S_{0}$ and $S_{1}$ are both zero, the same method is applied
to the second antiderivative $^{\left(2\right)}v(0,t)=\int_{-\infty}^{t}\int_{-\infty}^{u_{2}}v(0,u_{1})\,du_{1}du_{2}$,
the Fourier transform of which reads as $V(0,\omega)/\left(i\omega\right)^{2}$.
Denoting $S_{2}$ as the area of $^{\left(2\right)}v(0,t)$ and $t_{2}$
as its center of gravity \cite{re1}, we finally get the output signal
\begin{equation}
v(\ell,t)=S_{2}\,\ddot{h}(t-t_{2})\label{eq:quarantecinqquater}
\end{equation}
the amplitude of which scales as $\ell^{-3/2}$. This result applies
in particular to the precursors given by Eqs. (\ref{eq:trente}) and
(\ref{eq:quarantedeux}). The second antiderivative of the corresponding
input signals is such that $t_{2}=T_{c}/2$ in both cases, $S_{2}=2\pi/\omega_{c}^{3}$
in the first case and $S_{2}=2/\omega_{c}^{3}$ in the second case.
According to Eq. (\ref{eq:quarantecinqquater}), identical precursors
will be thus obtained by amplifying the input signal in the second
case by a factor $\pi$. This result is illustrated Fig. \ref{fig:cestpareil},
which spectacularly shows that quite different input signals can generate
the same precursor as long as they have the same integral properties.
Figure \ref{fig:cestpareil} has been drawn for a finite optical thickness
($\alpha_{c}\ell=12$) for which the precursors have sufficient
amplitude and can be really observed. Both are close to the asymptotic
form given by Eq. (\ref{eq:quarantecinqquater}) from which they become
undistinguishable for $\alpha_{c}\ell\geq30$.

\section{EFFECTS OF THE MEDIUM CONDUCTION\label{sec:Conduction}}

Debye media have always some conductivity. In the reference case of
water, this conductivity $\sigma$ is essentially static in the radiofrequency
and microwave domains \cite{ja98,ca09} and ranges from about $5\times10^{-6}\,\mathrm{S/m}$
for ultrapure water (only due to $H^{+}$ and $OH^{-}$ ions) to
$5\,\mathrm{S/m}$ for typical seawater (mainly due to dissolved
salts). In the presence of conductivity, the complex refractive index
reads as 
\begin{equation}
\tilde{n}(\omega)=\sqrt{n_{\infty}^{2}+\frac{n_{0}^{2}-n_{\infty}^{2}}{1+i\omega\tau}+\frac{\sigma}{i\omega\varepsilon_{0}}}\label{eq:quarantesix}
\end{equation}

When the conductivity is small enough, the main contribution to $\tilde{n}(\omega)$
and thus to $H(\ell,\omega)$ is expected to originate from frequencies
such that the last term in Eq. (\ref{eq:quarantesix}) may be treated
as a perturbation. Repeating the procedure that led to Eq. (\ref{eq:sept}),
we then get 
\begin{equation}
\ln\left[H\left(\ell,\omega\right)\right]\approx-i\omega t_{B}-\frac{\omega^{2}}{4\beta^{2}}-\frac{\sigma\ell Z_{0}}{2},\label{eq:quarantesept}
\end{equation}
where $Z_{0}=\sqrt{\mu_{0}/\left(n_{0}^{2}\varepsilon_{0}\right)}$
is the characteristic impedance of the medium at $\omega=0$ when
$\sigma=0$. Using again a time retarded by $t_{B}$, we finally get
the approximate transfer function $H_{a}(\ell,\omega)$ 
\begin{equation}
H_{a}(\ell,\omega)=\exp\left(-\frac{\sigma\ell Z_{0}}{2}\right)e^{-\frac{\omega^{2}}{4\beta^{2}}}\label{eq:quarantehuit}
\end{equation}
and the approximate impulse response $h_{a}\left(\ell,t\right)$ 
\begin{equation}
h_{a}\left(\ell,t\right)=\frac{\beta}{\sqrt{\pi}}\,e^{-\beta^{2}t^{2}}\exp\left(-\frac{\sigma\ell Z_{0}}{2}\right).\label{eq:quaranteneuf}
\end{equation}
At this order of approximation, the effect of the medium conductivity
is simply to reduce the impulse response and thus the\emph{ response
to any input signal }by the factor $F=\exp\left(\frac{\sigma\ell Z_{0}}{2}\right)$.
The effect of the medium conductivity will be negligible when the
product of its conductance $\sigma\ell$ by its characteristic impedance
$Z_{0}$ is small compared to $1$. For $\ell=1\,\mathrm{m}$, this
condition is practically realized with ordinary drinking water whose
typical conductivity ($10^{-3}\,\mathrm{S/m}$) leads to $F\approx1.02$.
The value $F=2$ is attained for $\sigma\approx0.033\,\mathrm{S/m}$
and numerical simulations shows that $h_{a}\left(\ell,t\right)$ is
then a very good approximation of the exact impulse response $h\left(\ell,t\right)$.
We, however, note that the area of $h_{a}\left(\ell,t\right)$ is only
$1/F$ whereas that of the exact impulse function remains equal to
$1$. The reason for this apparent discrepancy is that the low frequencies,
not fully taken into account in the previous calculation, originate
a very small but very long tail whose area is the missing area. This
tail is well visible on Fig. \ref{fig:h(t)s0,1=000026Canonique}, obtained
for $\sigma\approx0.1\,\mathrm{S/m}$. 
\begin{figure}
\begin{centering}
\includegraphics[width=85mm]{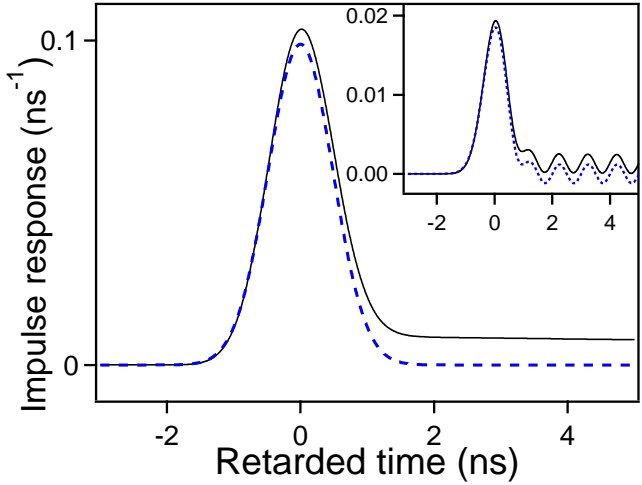} 
\par\end{centering}
\protect\caption{(Color online) Impulse response of water for $\ell=1\,\mathrm{m}$
($t_{B}\approx29.6\,\mathrm{ns}$) and $\sigma\approx0.1\,\mathrm{S/m}$.
The solid (dashed) line is the exact numerical result (our approximate
analytical result). Inset: corresponding response to the canonical
input signal $u_{H}(t)\sin\left(\omega_{c}t\right)$ with $\omega_{c}=2\pi\times10^{9}\,\mathrm{s}^{-1}$.\label{fig:h(t)s0,1=000026Canonique}}
\end{figure}
Though $F$ is now as large as $8.3$, Eq. (\ref{eq:quaranteneuf})
continue to provide a good approximation of the main part of the impulse
response. The inset of Fig. \ref{fig:h(t)s0,1=000026Canonique} shows
the output signal generated by the canonical input signal $u_{H}(t)\sin\left(\omega_{c}t\right)$
with $\omega_{c}=2\pi\times10^{9}\,\mathrm{s}^{-1}$ as previously.
As expected, $v(\ell,t)$ is fairly well reproduced by the analytical
expression $\mathrm{Im}\left[\tilde{v}(\ell,t)\right]/F$, where $\tilde{v}(\ell,t)$
is given by Eq. (\ref{eq:douze}), and the long tail of $h\left(\ell,t\right)$
originates the upshift of the main signal.

Before considering the general case, it is instructive to examine
the effects of the sole conductivity. This is achieved by neglecting
the second term under the square root in Eq. (\ref{eq:quarantesix}).
The transfer function is then reduced to $H_{c}(\ell,\omega)$ of
the form 
\begin{equation}
H_{c}(\ell,\omega)=\exp\left[-t_{\infty}\sqrt{\left(s+\gamma\right)^{2}-\gamma^{2}}\right],\label{eq:cinquante}
\end{equation}
where $s=i\omega$, $t_{\infty}=n_{\infty}\ell/c$ and $\gamma=\sigma/\left(2\varepsilon_{0}n_{\infty}^{2}\right)$.
Using the inverse Laplace transform of $\exp\left[-t_{\infty}\sqrt{s^{2}-\gamma^{2}}\right]$
as given in Ref \cite{ab72} and translating $s$ by $\gamma$, we get
the impulse response 
\begin{multline}
h_{c}\left(\ell,t\right)=\gamma t_{\infty}\frac{\mathrm{I_{1}}\left(\gamma\,\sqrt{t^{2}-t_{\infty}^{2}}\right)}{\sqrt{t^{2}-t_{\infty}^{2}}}e^{-\gamma t}u_{H}(t-t_{\infty})\\
+e^{-\gamma t_{\infty}}\delta\left(t-t_{\infty}\right).\label{eq:cinquanteetun}
\end{multline}
Here $\mathrm{I}_{1}(z)$ designates the first-order modified Bessel
function and $t$ is the real time. A more rigorous demonstration
of this result, showing its consistency with the boundary conditions,
can be found in Ref \cite{lo94}. For $t^{2}\gg t_{\infty}^{2}$ and $\gamma t\gg1$,
Eq. (\ref{eq:cinquanteetun}) takes the asymptotic form 
\begin{equation}
h_{c}\left(\ell,t\right)\approx\sqrt{\frac{\mu_{0}\sigma\ell^{2}}{4\pi t^{3}}}\exp\left(-\frac{\mu_{0}\sigma\ell^{2}}{4t}\right),\label{eq:cinquantedeux}
\end{equation}
which is maximum for $t=t_{m}=\mu_{0}\sigma\ell^{2}/6$ with an amplitude
\begin{equation}
h_{m}=\sqrt{\frac{54}{\pi e^{3}}}\left(\frac{1}{\mu_{0}\sigma\ell^{2}}\right).\label{eq:cinquanttrois}
\end{equation}

Coming back to the general problem, Eq. (\ref{eq:quaranteneuf}) is
expected to provide a good approximation of the precursor if, at the
time of its maximum, the ratio $R$ of $h_{a}\left(\ell,t\right)$
over $h_{c}\left(\ell,t\right)$is large. Using Eqs. (\ref{eq:quaranteneuf})
and (\ref{eq:cinquantedeux}) and taking into account that $n_{0}^{2}\gg n_{\infty}^{2}$,
we get the approximate expression 
\begin{equation}
R\approx n_{0}\sqrt{\frac{2\varepsilon_{0}}{\sigma\tau}}\exp\left(-\frac{Z_{0}\sigma\ell}{4}\right)=n_{0}\sqrt{\frac{2\varepsilon_{0}}{F\sigma\tau}}\label{eq:cinquantequatre}
\end{equation}
In the conditions of the inset of Fig. \ref{fig:h(t)s0,1=000026Canonique}
($\ell=1\,\mathrm{m}$ , $\sigma\approx0.1\,\mathrm{S/m}$.), $R$
was about $14$ and Eq.(\ref{eq:quaranteneuf}) actually provided
a good approximation of the medium response to the canonical input
signal with $\omega_{c}=2\pi\times10^{9}\,\mathrm{s}^{-1}$ . 
\begin{figure}
\begin{centering}
\includegraphics[width=85mm]{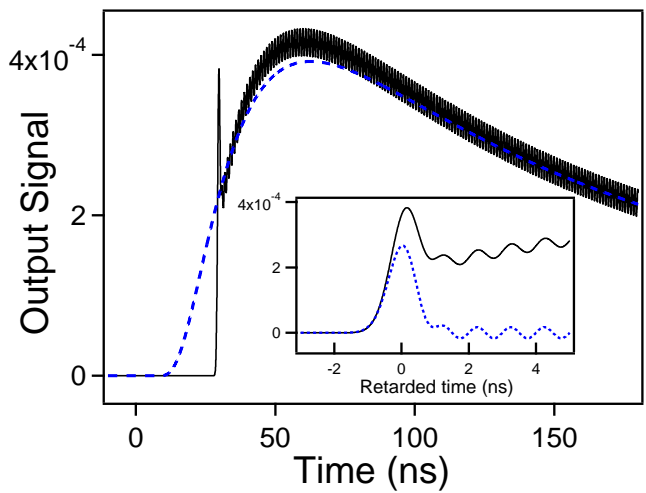} 
\par\end{centering}
\protect\caption{(Color online) Response to the canonical input signal for $\sigma\approx0.3\,\mathrm{S/m}$
(other parameters as for the inset of Fig. \ref{fig:h(t)s0,1=000026Canonique}).
The solid (dashed) line is the exact numerical result (the analytical
result obtained by considering the sole effects of conductivity).
Inset: enlargement of the output signal around $t=t_{B}$. Here the
dashed line is the analytical result derived from Eq. (\ref{eq:quaranteneuf}).\label{fig:canonique_s=00003D00003D00003D0,3}}
\end{figure}
Figure \ref{fig:canonique_s=00003D00003D00003D0,3} shows the result
obtained for $\sigma\approx0.3\,\mathrm{S/m}$, leading to $R\approx1$.
The response obtained by considering the sole effects of conductivity
\cite{re2} approximates fairly well the mean value of the exact response
for $t>t_{B}$. The coincidence becomes exact for $t\gg t_{B}$, the
long-time behavior being mainly determined by the low frequencies
where the conductivity term prevails on the polarization term in Eq. (\ref{eq:quarantesix}).
On the other hand, the inset shows that the beginning of the response
is perfectly reproduced by the analytical response derived from Eq. (\ref{eq:quaranteneuf}).
Main signal and precursor obviously decrease with the conductivity.
For $\sigma\approx0.4\,\mathrm{S/m}$, the former remains visible
whereas the latter appears as a slight overshot of amplitude about
$10^{5}$ times smaller than that of the input signal and would be
probably undetectable in a real experiment. 
\begin{figure}
\begin{centering}
\includegraphics[width=85mm]{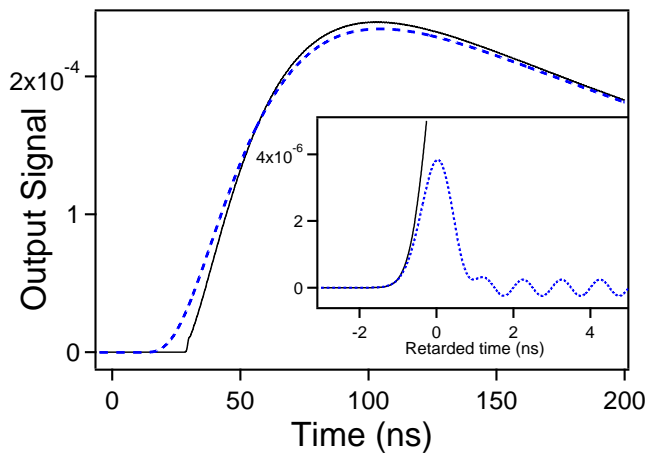} 
\par\end{centering}
\protect\caption{(Color online) Same as Fig. \ref{fig:canonique_s=00003D00003D00003D0,3}
for $\sigma\approx0.5\,\mathrm{S/m}$. Note that the scales of the
inset and the main figure are very different.\label{fig:canonique_s=00003D00003D00003D0,5}}
\end{figure}
Figure \ref{fig:canonique_s=00003D00003D00003D0,5} shows that main
signal and precursor become practically invisible for $\sigma\approx0.5\,\mathrm{S/m}$.
The overall response is well fitted by the analytical function derived
from Eq. (\ref{eq:cinquanteetun}) \cite{re2} but Eq. (\ref{eq:quaranteneuf})
continue to perfectly reproduce the very first beginning of the signal
(inset). Finally, as the conductivity increases, the fit
of the output signal by $h_{c}\left(\ell,t\right)/\omega_{c}$ improvies. Any
trace of precursor then disappears and the response is reduced to
a broad signal of amplitude (duration) scaling as $1/\ell^{2}$ (as
$\ell^{2}$). See Eq. (\ref{eq:cinquantedeux}) and (\ref{eq:cinquanttrois}).

The simulations of Figs. \ref{fig:h(t)s0,1=000026Canonique} (inset),
\ref{fig:canonique_s=00003D00003D00003D0,3}, and \ref{fig:canonique_s=00003D00003D00003D0,5}
have been made for $\ell=1\,\mathrm{m}$ and $\omega_{c}=2\pi\times10^{9}\,\mathrm{s}^{-1}$.
These parameters are those of a realistic experiment on water and,
such that the Brillouin precursor generated by the canonical input
signal predominates on the main field which remains visible.
The analytical results are, however, general. They notably show that
the validity domains of the low- and high-conductivity approximations
are not entirely determined by the absolute value of the conductivity
as considered in Ref \cite{ca09} but strongly depend on the propagation
length. For example, a conductivity $\sigma\approx3.3\times10^{-2}\,\mathrm{S/m}$
suffices to be perfectly inside the low conductivity domain ($R\approx51$)
when $\ell=1\,\mathrm{m}$, whereas a conductivity as low as $\sigma\approx1.72\times10^{-3}\,\mathrm{S/m}$
is required to attain the same value of $R$ when $\ell=100\,\mathrm{m}$
. Note additionally that the amplitude-reduction factor $F$ is as
large as $38$ in the latter case but $2$ in the former.

\section{CONCLUSION\label{sec:CONCLUSION}}

Sommerfeld and Brillouin precursors originate from interrelated effects
of group-velocity dispersion and of frequency-dependent absorption
of the medium. The former prevail in the formation of the Sommerfeld
precursors \cite{som14,fa03,bm12} but both effects generally contribute
to that of the Brillouin precursors \cite{bri14,bri32,bm12}. In Debye
media, however, as soon as the Brillouin precursor is distinguishable
from the main field, we have shown that the impulse response of the
medium is perfectly fitted by the Gaussian obtained by neglecting
the dispersion effects. By means of simple convolutions, we have then
obtained explicit analytical expressions of the response of the medium
to a wide class of input signals and extracted from them those of
the precursor. Our main results are summarized below.

For input signals with a single initial discontinuity, the Brillouin
precursor is well approximated by a Gaussian (Figs. \ref{fig:PrecursorAnalyticVsApprSin}
and \ref{fig:Precusor_Cos-vs-FIT}) or a Gaussian derivative (Figs. \ref{fig:PrecusorAnalyticVsApprCos}
and \ref{fig:Precusor_Sin-vs-FIT}), depending on the parity of
the discontinuity order. This approximation is excellent in the case
of the canonical input signal $u_{H}(t)\sin\left(\omega_{c}t\right)$
considered by Sommerfeld and Brillouin (Fig. \ref{fig:PrecursorAnalyticVsApprSin})
and an exact expression of the precursor amplitude is obtained in
this case {[}Eq. (\ref{eq:seize}){]}. When the input signal has two
successive discontinuities separated by one period of the carrier
frequency, the quasi-destructive interference of the corresponding
precursors may generate a unique precursor with a second-Gaussian-derivative
shape (Figs. \ref{fig:OneperiodeRectangularCosPulse}, \ref{fig:LinearPrecusorTr=00003D00003D00003D1=0000260,5Tc}a,
and \ref{fig:cestpareil}).

In the limit where the optical thickness of the medium at the carrier
frequency is very large, the above-mentioned shapes of precursor are
exact and the precursors are simply proportional to the impulse response,
to its first or to its second derivative, with amplitude scaling with
the propagation distance $\ell$ as $\ell^{-1/2}$, $\ell^{-1}$ and
$\ell^{-3/2}$ respectively. These remarkable asymptotic properties
are not specific to particular input signals but are common to the
precursors generated by all the input signals having the same integral
properties. The same precursor can thus be generated by quite different
input signals (Fig. \ref{fig:cestpareil}).

An eventual static conductivity of the medium does not affect significantly
the shape of the output signal and in particular of the precursor
when the parameter $R$ given Eq.(\ref{eq:cinquantequatre}) is large
(low conductivity limit). The main effect of the conductivity is then
an overall reduction of the medium response {[}see Eq. (\ref{eq:quaranteneuf})
and Fig. \ref{fig:h(t)s0,1=000026Canonique}{]}. In the opposite case
where $R\ll1$ (high conductivity limit), the precursor disappears
and the output signal is well approximated by the broad signal obtained
by neglecting the polarization contribution, the amplitude of which
scales as $\ell^{-2}$ {[}see Eq. (\ref{eq:cinquanttrois}) and Fig. \ref{fig:canonique_s=00003D00003D00003D0,5}{]}.

The results obtained in our study cover various
situations. Only the Brillouin precursor generated
by the canonical input signal has been actually evidenced in real
experiments \cite{sto01}. Our theoretical analysis is expected to
stimulate complementary experimental works.

\section*{ACKNOWLEDGEMENTS }

This work has been partially supported by Ministry of Higher Education
and Research, Nord-Pas de Calais Regional Council and European Regional
Development Fund (ERDF) through the Contrat de Projets \'{E}tat-R\'{e}gion
(CPER) 2007\textendash 2013, as well as by the Agence Nationale de
la Recherche through the LABEX CEMPI project (ANR-11-LABX-0007).


\begin{thebibliography}{0}
\expandafter\ifx\csname natexlab\endcsname\relax\def\natexlab#1{#1}\fi
\expandafter\ifx\csname bibnamefont\endcsname\relax
  \def\bibnamefont#1{#1}\fi
\expandafter\ifx\csname bibfnamefont\endcsname\relax
  \def\bibfnamefont#1{#1}\fi
\expandafter\ifx\csname citenamefont\endcsname\relax
  \def\citenamefont#1{#1}\fi
\expandafter\ifx\csname url\endcsname\relax
  \def\url#1{\texttt{#1}}\fi
\expandafter\ifx\csname urlprefix\endcsname\relax\def\urlprefix{URL }\fi
\providecommand{\bibinfo}[2]{#2}
\providecommand{\eprint}[2][]{\url{#2}}

\end{thebibliography}


\begin{thebibliography}{10}
\bibitem{som14} A. Sommerfeld, Ann. Phys. (Leipzig) \textbf{44},
177 (1914).

\bibitem{bri14} L. Brillouin, Ann. Phys. (Leipzig) \textbf{44}, 203
(1914).

\bibitem{bri32}L. Brillouin, \emph{Comptes Rendus du Congr\`{e}s International
d'Electricit\'{e}, Paris 1932} (Gauthier-Villars, Paris 1933), Vol.2, pp. 739-788.

\bibitem{bri60} L. Brillouin,\emph{Wave Propagation and Group Velocity}
(Academic Press, New York, 1960). Authorized translations in English
of Ref \cite{som14,bri14,bri32} can be found in this book.

\bibitem{ou09} K.E. Oughstun, \emph{Electromagnetic and Optical Pulse
Propagation 2 : Temporal Pulse Dynamics in Dispersive Attenuative
Media} (Springer, New York, 2009).

\bibitem{je09} H. Jeong, U.L. Österberg, and T. Hansson, J. Opt.
Soc. Am. B \textbf{26}, 2455 (2009).

\bibitem{ou10} K.E. Oughstun, N.A. Cartwright, D.J. Gauthier, and
H. Jeong, J. Opt. Soc. Am. B \textbf{27}, 1664 (2010).

\bibitem{bm11} B. Macke and B. S\'{e}gard, J. Opt. Soc. Am. B \textbf{28},
450 (2011).

\bibitem{cia11} A. Ciarkowski, Int. J. Electron. Telecommun. \textbf{57},
251 (2011).

\bibitem{bm12} B. Macke and B. S\'{e}gard, Phys. Rev. A \textbf{86},
013837 (2012).

\bibitem{pl69} P. Pleshko and I. Palocz, Phys. Rev. Lett. \textbf{22},
1201 (1969).

\bibitem{aa88} J. Aaviksoo, J. Lippmaa, and J. Kuhl, J. Opt. Soc.
Am. B \textbf{5}, 1631 (1988).

\bibitem{bm13} B. Macke and B. S\'{e}gard, Phys. Rev. A \textbf{87},
043830 (2013).

\bibitem{bs87} B. S\'{e}gard, J. Zemmouri, and B. Macke, Europhys. Lett.
\textbf{4}, 47 (1987). See in particular Fig. 2 in this reference.

\bibitem{aa91} J. Aaviksoo, J. Kuhl, and K. Ploog, Phys. Rev. A \textbf{44},
5353(R) (1991).

\bibitem{je06} H. Jeong, A. M. C. Dawes, and D. J. Gauthier, Phys.
Rev. Lett. \textbf{96}, 143901 (2006).

\bibitem{wei09} D. Wei, J.F. Chen, M.M.T. Loy, G.K.L. Wong, and S.
Du, Phys. Rev. Lett. \textbf{103}, 093602 (2009).

\bibitem{bm10} B. Macke and B. S\'{e}gard, Phys. Rev. A \textbf{81},
015803 (2010). See Fig. 1 in this paper.

\bibitem{zh13} Z.Q. Zhou, C.F. Li, and G.C. Guo, Phys. Rev. A \textbf{87},
045801 (2013).

\bibitem{fa03} E. Falcon, C. Laroche, and S. Fauve, Phys. Rev. Lett.
\textbf{91}, 064502 (2003).

\bibitem{de30} P. Debye, \emph{Polar Molecules} (Dover, New York
1929).

\bibitem{al89}R. Albanese, J. Penn, and R. Medina, J. Opt. Soc. Am.
A \textbf{6}, 1441 (1989).

\bibitem{ro96}T.M. Roberts and P.G. Petropoulos, J. Opt. Soc. Am.
A \textbf{13}, 1204 (1996).

\bibitem{fa97}E.G. Farr and C.A. Frost, U.S Air Force, Laboratory Technical Report No. WL-TR-1997-7050, 1997. Available at http://www.dtic.mil/dtic/tr/fulltext/u2/a328788.pdf.

\bibitem{ka98} A. Karlsson and S. Rike, J. Opt. Soc. Am. A\textbf{
15}, 487 (1998).

\bibitem{li01}Y. Liu and W. Wang, IEEE Trans. Electromagn. Compat.
\textbf{43}, 223 (2001).

\bibitem{sto01}D. C. Stoudt, F. E. Peterkin, and B. J. Hankla, NSWC
Report No. JPOSTC-CRF-005-03, 2001. Available at http://ece-research.unm.edu/summa/notes/In/IN622.pdf.

\bibitem{ro04}T.M. Roberts, IEEE Trans. Antennas Propag. \textbf{52},
310 (2004).

\bibitem{ou05}K.E. Oughstun, IEEE Trans. Antennas Propag. \textbf{53},
1582 (2005).

\bibitem{pi09}M. Pieraccini, A. Bicci, D. Mecatti, G. Macaluso, and
C. Atzeni, IEEE Trans. Antennas Propag. \textbf{\noun{57}}, 3612 (2009).

\bibitem{sa09}R. Safian, C.D. Sarris, and M. Mojahedi, IEEE Trans.
Antennas Propag. \textbf{57}, 3676 (2009).

\bibitem{da10}M. Dawood, H.U.R. Mohammed, and A.V. Alejos, Electron. Lett. \textbf{46}, 1645 (2010).

\bibitem{ca11}N. Cartwright, IEEE Trans. Antennas Propag. \textbf{59},
1571 (2011).

\bibitem{ou12}K.E. Oughstun and C.L. Palombini, in \emph{Proceedings of the
International Microwave Symposium Digest (MTT), 2012 IEEE MTT-S International} (IEEE, New York, 2012), pp. 1-3.

\bibitem{pap87}We use the definitions, sign conventions, and results
of the linear system theory. See, for example, A.Papoulis, \emph{The
Fourier Integral and Its Applications} (Mc Graw Hill, New York, 1987).

\bibitem{se81}D. Segelstein,  M.S. Thesis, University of Missouri, Kansas City 1981 (unpublished). For
a direct access to the relevant numerical data, see: http://www.philiplaven.com/Segelstein.txt

\bibitem{bu04}N.S. Bukhman, Quantum Electron. \textbf{34}, 299 (2004).

\bibitem{re0} Causality implies that $h(\ell,t)=0$ for real time
less than $n_{\infty}\ell/c$, that is for retarded time less than
$(n_{\infty}-n_{0})\ell/c$.

\bibitem{ni10}\emph{NIST Handbook of Mathematical functions}, edited
by F.W.J. Olver, D.W. Lozier, R.F. Boisvert, and C.W. Clark (Cambridge University Press, Cambridge, 2010).

\bibitem{re1} The area (the center of gravity) is simply given by
the first non-zero term (the following term) of the expansion of $V(0,\omega)$
in power series of $i\omega$.

\bibitem{ja98}J.A. Jackson, \emph{Classical Electrodynamics}, $3rd$
ed. (Wiley, New York, 1998).

\bibitem{ca09}N.A. Cartwright and K.E. Oughstun, \emph{Antennas and Propagation
Society International Symposium, 2009. APSURSI '09. IEEE} (IEEE, Charleston, 2009) pp. 1-4.

\bibitem{ab72}\emph{Handbook of Mathematical Functions}, edited by
M. Abramowitz and I. A. Stegun (Dover, New York, 1972). See Eq.(29.3.96),
p. 1027, in this book.

\bibitem{lo94}J. LoVetri and J.B. Ehrman, IEEE Trans. Electromagn.
Compat. \textbf{36}, 221 (1994). See Eq.(25) in this article, and the references cited therein.

\bibitem{re2}$h_{c}(\ell,t)$ evolving slowly at the scale of the
period $T_{c}=2\pi/\omega_{c}$, this response is simply $S_{0}h_{c}(\ell,t-t_{0})$
with $S_{0}=1/\omega_{c}$ and $t_{0}=0$. See Eq. (\ref{eq:quarantecinqbis})
and Ref \cite{re1}. Note also that the location and the amplitude of
the maximum are in excellent agreement with those derived from the
asymptotic form of $h_{c}(\ell,t)$ .\end{thebibliography}
\end{document}